\documentclass{elsart4-1}

 \usepackage{graphicx}

\usepackage{amssymb}
\usepackage{amsmath}

\usepackage[english,francais]{babel}

\newtheorem{e-proposition}[theorem]{Proposition}

\newtheorem{e-definition}[theorem]{Definition\rm}

\newcommand{\eqname}[1]{\label{eq:#1}}
\newcommand{\bgar}{\begin{eqnarray}}
\newcommand{\enar}[1]{\label{eq:#1}\end{eqnarray}}

\newcommand{\expect}[1]{\left\langle #1 \right\rangle}

\newcommand{\eq}[1]{(\ref{eq:#1})}

\newcommand{\Psihd}{\hat\Psi^\dagger}

\newcommand{\Psih}{\hat\Psi}

\newcommand{\bhd}{\hat b^\dagger}
\newcommand{\bh}{\hat b}

\def\og{\leavevmode\raise.3ex\hbox{$\scriptscriptstyle\langle\!\langle$~}}
\def\fg{\leavevmode\raise.3ex\hbox{~$\!\scriptscriptstyle\,\rangle\!\rangle$}}

\begin{document}

\begin{frontmatter}


\selectlanguage{english}
\title{Superfluidity of the 1D Bose gas}

\vspace{-2.6cm}

\selectlanguage{francais}
\title{Superfluidit\'e du gaz de Bose 1D}


\selectlanguage{english}

\author[LKB,TN]{Iacopo Carusotto}
\ead{Iacopo.Carusotto@lkb.ens.fr}
\author[LKB]{Yvan Castin}
\ead{Yvan.Castin@lkb.ens.fr}


\address[LKB]{Laboratoire Kastler Brossel, \'Ecole Normale
Sup\'erieure, 24 rue Lhomond, 75231 Paris Cedex 05, France}
\address[TN]{BEC-INFM Research and Development Center, Trento, Italy}


\begin{abstract}
We have investigated the superfluid properties of a ring of weakly interacting
and degenerate 1D Bose gas at thermal equilibrium with a rotating vessel.
The conventional definition of superfluidity
predicts that the gas has a significant superfluid fraction only in the
Bose condensed regime. In the opposite regime, it is found that a superfluid behaviour
can still be identified when the probability distribution of the total momentum
of the gas has a multi-peaked structure, revealing unambiguously the existence of
`superfluid' supercurrent states that did not show up
in the conventional definition of superfluidity.

\vskip 0.5\baselineskip

\selectlanguage{francais}
\noindent{\bf R\'esum\'e}
\vskip 0.5\baselineskip
\noindent
Cet article \'etudie, dans le r\'egime d'interaction faible, les
propri\'et\'es superfluides d'un gaz de Bose unidimensionnel
confin\'e sur un anneau et \`a l'\'equilibre thermodynamique dans un r\'ef\'erentiel tournant.
La d\'efinition habituelle de la superfluidit\'e pr\'edit que ce gaz a
une fraction superfluide appr\'eciable seulement s'il est aussi un
condensat de Bose-Einstein. Dans le r\'egime non condens\'e,
nous trouvons cependant qu'il est possible d'identifier un comportement superfluide
en consid\'erant la distribution de probabilit\'e de l'impulsion totale
du gaz~: il existe un r\'egime o\`u cette distribution comporte plusieurs pics bien
s\'epar\'es, ce qui d\'emontre l'existence de super-courants superfluides qui passent
inaper\c{c}us dans la d\'efinition habituelle de la superfluidit\'e.

\keyword{Superfluidity; one-dimensional; Bose gas}
\vskip 0.5\baselineskip
\noindent{\small{\it Mots-cl\'es~:} Superfluidit\'e; unidimensionnel;
  gaz de Bose}}
\end{abstract}
\end{frontmatter}

\selectlanguage{english}
\section{Introduction}
\label{sec:Intro}
Is the weakly interacting 1D Bose gas superfluid~?

Diverging answers can be found in the literature.
The Landau criterion~\cite{Huang} gives a positive answer, since the dispersion
of elementary excitations is linear at low momenta.
In the thermodynamic limit, it is argued in \cite{Popov} that,
on the contrary, the 1D Bose gas cannot be superfluid at finite temperature:
it does not experience any phase transition, and the field correlation function
vanishes exponentially at large distances, rather than with a power law.
A calculation based on the Bethe ansatz, with some additional assumptions on the
accessible many-body states, concludes that superfluidity is possible
\cite{Sonin}. As shown in \cite{Leggett,Svistunov} one of the subtleties of the issue
is that there are actually different definitions of superfluidity, one
of them based on a static (that is thermal equilibrium) property of the system,
the other one involving a dynamical response of the system. In 1D, these
two definitions are found to dramatically differ in the thermodynamic limit
\cite{Svistunov}.

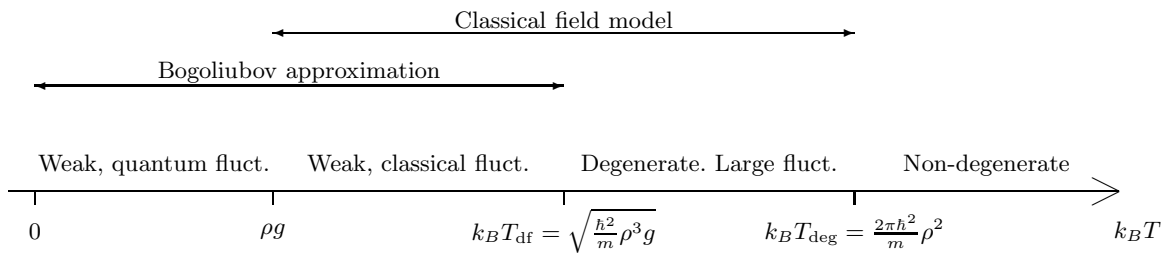
\begin{figure}[htb]
\begin{center}
\begin{picture}(430,150)(0,50)
\put(0,100){\line(1,0){420}}
\put(420,100){\line(-2,-1){10}}
\put(420,100){\line(-2,1){10}}
\put(427,85){\makebox(0,0){$k_B T$}}
\put(10,85){\makebox(0,0){$0$}}
\put(10,95){\line(0,1){5}}

\put(100,85){\makebox(0,0){$\rho g$}}
\put(100,95){\line(0,1){5}}

\put(210,85){\makebox(0,0){$k_B T_{\rm df}=\sqrt{\frac{\hbar^2}{m}\rho^3 g}$}}
\put(210,95){\line(0,1){5}}

\put(320,85){\makebox(0,0){$k_B T_{\rm deg}=\frac{2\pi\hbar^2}{m}\rho^2$}}
\put(320,95){\line(0,1){5}}

\put(55,110){\makebox(0,0){Weak, quantum fluct.}}
\put(155,110){\makebox(0,0){Weak, classical fluct.}}
\put(265,110){\makebox(0,0){Degenerate. Large fluct.}}
\put(370,110){\makebox(0,0){Non-degenerate}}

\put(10,140){\vector(1,0){200}}
\put(210,140){\vector(-1,0){200}}
\put(110,145){\makebox(0,0){Bogoliubov approximation}}

\put(100,160){\vector(1,0){220}}
\put(320,160){\vector(-1,0){220}}
\put(210,165){\makebox(0,0){Classical field model}}

\end{picture}
\caption{\label{fig:T_diagram} Scheme of the different approaches used in this paper and
  the corresponding applicability domains. The three temperature
  scales on the axis differ by a factor $(\hbar \rho/ mg)^{1/2}\gg 1$
  in the weakly interacting regime. `fluct.' stands for `density fluctuations'.}
\end{center}
\end{figure}

In this paper, we restrict to the strict thermal equilibrium regime, in a case where
the gas can exchange momentum with a rotating vessel with walls that are smooth, at least at
the macroscopic scale.  Note that this differs from the usual stirring procedures used in experiments with
condensates, where a macroscopic rotating defect is applied \cite{Dalibard,Ketterle,JILA,Foot}.
We investigate the superfluid properties of the quantum gas in various
limiting cases, from the ideal Bose gas to the weakly interacting Bose
gas with weak density fluctuations, where the Bogoliubov approximation applies. 
We also consider an exactly solvable
classical field model that allows also to study the interacting case with large density fluctuations.
The applicability domain of this model has some overlap with the one of the Bogoliubov
approximation, see fig.\ref{fig:T_diagram}, and in this overlap domain the two approaches
give coincident results.
Investigations are performed by considering not only the mean momentum of the rotating gas, but
also the whole probability distribution of the total momentum: this allows to reach a much deeper
physical understanding of the static aspects of the problem.

\section{General considerations}

\subsection{The physical model}
\label{sec:model}

Consider a one-dimensional Bose gas as described
in a second-quantization approach by the Hamiltonian:
\begin{equation}
\label{eq:Hamilt}
{\mathcal
  H}_0=-\frac{\hbar^2}{2m}\int_0^L\!dx\,\Psihd(x)\frac{\partial^2}{\partial x^2}
  \Psih(x)+\frac{g}{2}
\int\!dx\,\Psihd(x)\Psihd(x)\Psih(x)\Psih(x).
\end{equation}
The $\Psih(x)$ and $\Psihd(x)$ operators are respectively the
destruction and creation operators for a boson at point $x$. They obey
standard bosonic commutation rules
$[\Psih(x),\Psihd(x')]=\delta(x-x')$.
The spatial coordinate $x$ runs on a ring of length $L$ with periodic
boundary conditions, $m$ is the atomic mass, and the strength of
local interactions is quantified by $g$. We shall restrict in this paper
to the repulsive and weakly interacting case so that we impose
$0<g\ll\hbar^2 \rho/m$~\cite{WeaklyInter1D} where $\rho$ is the mean density.

The Hamiltonian \eq{Hamilt} is a good description of a Bose gas in a
cylindrically symmetric toroidal trap~\cite{torus} provided (i) the transverse trapping
frequencies in the torus are much larger than both the temperature
and the interaction energy per particle, $\hbar\omega_{\rho,z}\gg k_B
T, \rho g$, and (ii) the radius $L/2\pi$ of the torus is much larger than the width
of the transverse harmonic oscillator 
ground state which allows to neglect curvature effects in the kinetic
energy. In this regime, the system is effectively 1D with periodic boundary
conditions.

The gas is assumed to be at thermal equilibrium in an uniformly
rotating frame, which mimics the presence of a rotating vessel
containing the fluid; transfer of angular momentum from the vessel to
the fluid is assumed to be possible, so that thermal equilibrium between the
two can be attained. Rotation is then described by the following
additional term in the Hamiltonian:
\begin{equation}
\eqname{rhoRot}
{\mathcal H}_{\rm rot}=-v_{\rm rot} P,
\end{equation}
where $v_{\rm rot}=\Omega L/2\pi$ is the rotation velocity and $P$ is
the total momentum operator of the gas:
\begin{equation}
\eqname{Poper}
P=-i\hbar\int\!dx\,\Psihd(x)\frac{\partial}{\partial x}\Psih(x).
\end{equation}
Notice that the total momentum operator $P$ gives the total mass
current in the laboratory frame~\cite{CCT2001-2}.

\subsection{Definition of the normal fraction of the gas}
We define the generalized normal fraction $f_n$ of the gas as:
\begin{equation}
f_n=\frac{\expect{P}}{{N}\,m\,v_{\rm rot}},
\eqname{fn}
\end{equation}
where the expectation value of the total momentum operator is taken in
thermal equilibrium and $N$ is the total number of
particles. The usual normal fraction of the gas is the limit $f_n^0$
of $f_n$ for $v_{\rm rot}\rightarrow 0$, and the corresponding superfluid
fraction is $1-f_n^0$ \cite{Leggett}.
For a rigid body $f_n=1$, which means that it is at rest in the
rotating frame: the fluid is completely dragged by the walls of the
rotating vessel.
On the other hand $f_n=0$ for a pure superfluid: even if the
vessel is rotating, the fluid remains at rest
in the laboratory frame.

In the following sections, we
shall study in detail the behaviour of $f_{n}$ for a
weakly interacting one-dimensional Bose gas in different
temperature and density regimes. The temperature is assumed to be
always much larger than the spacing of single particle levels:
\begin{equation}
T\gg\frac{2\pi^2 \hbar^2}{m L^2},
\eqname{T>spac}
\end{equation}
that is the size $L$ of the system is assumed to be much longer than
the thermal de Broglie wavelength $\lambda$:
\begin{equation}
L\gg\lambda=\sqrt{\frac{2 \pi \hbar^2}{m k_B T}}.
\eqname{T>spac2}
\end{equation}

For our finite-size system, only velocity boosts $v$ which are integer
multiples of the characteristic velocity:
\begin{equation}
\eqname{v1}
v_1=\frac{2\pi \hbar}{mL}
\end{equation}
are allowed
\footnote{The effect a velocity boost of velocity $v$ on the many-body
wavefunction $\psi(x_1,\ldots,x_N)$ is
\begin{equation}
\psi'(x_1,\ldots,x_N)=\psi(x_1,\ldots,x_N)\,e^{i m v \sum_i
x_i\;/\hbar}.
\eqname{boost}
\end{equation}
Periodic boundary conditions therefore impose $m v =2 \pi \hbar M/L$, $M$
being an integer.}.
Galilean invariance under such boosts implies that if $\psi$ is an
eigenstate of \eq{Hamilt} of
energy $E$ and momentum $P$, the boosted state $\psi'$ is also an
eigenstate of energy $E'=E+vP+Nmv^2/2$ and momentum
$P'=P+Nmv$.
Provided the mean number of particles $N$ is kept constant,
the function $P(v_{\rm rot})$ giving the mean momentum $\expect{P}$ as a
function of velocity $v_{\rm rot}$ therefore satisfies a periodicity condition
of the form:
\begin{equation}
\eqname{periodicP}
P(v_{\rm rot}+v_1)=P(v_{\rm rot})+N m v_1.
\end{equation}
Because of the symmetry under spatial inversion, $P(v_{\rm rot})$ is an
odd function of $v_{\rm rot}$. This property, combined with \eq{periodicP},
implies that $P(v_1/2)=m N v_1/2$ and therefore $f_n(v_1/2)=1$.
For this reason, in the following we shall restrict the definition (\ref{eq:fn})
of the generalized normal fraction to the velocity range $v_{\rm rot}\in [-v_1/2,v_1/2]$.

\subsection{Probability distribution of the total momentum $P$}

In the previous subsection we have introduced the concept of normal
fraction $f_n$ of the gas in terms of the expectation value of the
total momentum operator $P$.
We shall extend our analysis by considering not only the
average value of $P$, but rather the complete probability distribution
$p(P)$, which gives the probability for the total momentum to be equal
to some given value $P$.
Notice that this probability distribution $p(P)$ is totally different
and distinct from the usual momentum distribution, which gives instead
the mean number of particles in each momentum state.

As usual, the first step for the determination of $p(P)$ is the
calculation of the corresponding characteristic function $g(\zeta)$:
\begin{equation}
\eqname{g_P}
g(\zeta)=\big\langle{e^{i\zeta P}}\big\rangle.
\end{equation}
The probability distribution $p(P)$ is then obtained as the Fourier
transform of $g$:
\begin{equation}
\eqname{p_P}
p(P)=\int\!\frac{d\zeta }{2\pi} e^{-i\zeta P}\,g(\zeta ).
\end{equation}

The slow rotation $v_{\rm rot}\rightarrow 0$ limit of the normal
fraction can be related to the variance of $p(P)$ in the
non-rotating $v_{\rm rot}=0$ system.
As the total momentum $P$ commutes with the Hamiltonian ${\mathcal
  H}_0$, the density matrix
of the fluid  in the canonical ensemble at a given inverse
temperature $\beta=1/k_B T$ can be
expanded for small $v_{\rm rot}$ as:
\begin{equation}
\eqname{RhoSmallV}
\rho=e^{-\beta({\mathcal H}_0-P v_{\rm rot})}\simeq
e^{-\beta{\mathcal H}_0}\big(1+\beta v_{\rm rot} P\big).
\end{equation}
This expression can be used to calculate the expectation value of the
momentum $P$ and then the small velocity limit $f_n^0$ of the normal
fraction.
\begin{equation}
f_n^0=\lim_{v_{\rm rot}\rightarrow 0} f_n=\frac{\expect{P^2}}{m\,k_B\,T\,{N}}.
\eqname{fn0}
\end{equation}
This relation also holds in the grand-canonical ensemble, as $P$ commutes with the number
of particles and as the chemical potential $\mu$ varies only to second order in $v_{\rm rot}$
for a fixed mean number of particles.  It is easy to see that for a Boltzmann gas of $N$ distinguishable and
non-interacting particles, the equipartition theorem of classical
statistical mechanics implies that
\begin{equation}
\expect{P^2}=\sum_i\expect{p_i^2}=m N k_B T,
\end{equation}
where $p_i$ is the momentum of the $i$-th particle. The system is therefore
totally normal $f_n^0=1$.

\section{Non-interacting gas}

The present section is devoted to a study of the rotational properties
of a non-interacting gas. In the grand-canonical ensemble, the
population of each one-particle state is described by the usual
Bose distribution:
\begin{equation}
n_k=\frac{1}{e^{\beta(\epsilon_k-\mu)}-1},
\eqname{Bose}
\end{equation}
where the momentum $k$ is quantized as usual as:
\begin{equation}
k=\frac{2\pi\hbar}{L}n,
\end{equation}
$n$ being an integer, and the single-particle energy
$\epsilon_k(v_{\rm rot})$ in
the rotating frame at $v_{\rm rot}$ is equal to:
\begin{equation}
\epsilon_k(v_{\rm rot})=\frac{\hbar^2 k^2}{2m}-\hbar k v_{\rm rot}.
\eqname{epsilonk}
\end{equation}
The mean number of particles is
\begin{equation}
{N}=\sum_k n_k,
\eqname{Nmean}
\end{equation}
and the mean momentum is
\begin{equation}
\expect{P}=\sum_k \hbar k \, n_k.
\eqname{Pmean}
\end{equation}
The normal fraction $f_n$ is then immediately obtained from its
definition \eq{fn}. Its zero-velocity value $f_n^0$ could also be
obtained from \eq{fn0} by taking into account the fact that for an
ideal gas one has:
\begin{equation}
\expect{n_k^2}=2\expect{n_k}^2+\expect{n_k}.
\eqname{Pmean2}
\end{equation}
As for $v_{\rm rot}=0$ a Bose-Einstein condensate can only appear in
the $k=0$ mode, the prediction for $f_n^0$ is not affected by the
non-physical grand-canonical fluctuations in the condensate mode.

\subsection{Non degenerate gas}

In the limit $\beta\mu\rightarrow -\infty$, the occupation of all single
particle modes is much smaller than unity and the Bose distribution \eq{Bose}
can be approximated by a Maxwell-Boltzmann law of the form:
\begin{equation}
n_k=e^{-\beta(\epsilon_k-\mu)}.
\eqname{nondeg}
\end{equation}
As $n_k$ is a slowly varying function of $k$ (the condition
\eq{T>spac} is always assumed), one can replace the
sum over $k$  in \eq{Nmean} and \eq{Pmean} by integrals
$\frac{L}{2\pi}\int\!dk$.
By switching
to the integration
variable $k'=k+mv/\hbar$, it is then immediate to see that:
\begin{equation}
\expect{P}=m {N}  v_{\rm rot},
\eqname{f_n1}
\end{equation}
which means that the fluid is completely normal and no superfluid
fraction is present. Notice that this conclusion does not depend on
the specific choice of the Maxwell-Boltzmann distribution \eq{nondeg}.
It rather depends on the facts that (i) the occupation number of the mode of
wavevector $k$ is a function of $\hbar k-m v_{\rm rot}$ only,
and (ii) sums can be replaced by integrals. Notice that the condition (i)
no longer holds
in the presence of interactions, which allows e.g.\ for superfluid
behaviour of the 3D Bose gas even in the thermodynamical limit.
For the ideal Bose gas, as we shall see in the next subsection, condition
(ii) is violated in the Bose-condensed regime.

\subsection{Classical field approximation}

If one assumes that the
temperature is larger than the absolute value of the
chemical potential $k_B T\gg|\mu|$, which corresponds to the limit $\beta\mu \rightarrow 0^-$,
the Bose distribution \eq{Bose} can be
approximated by the classical field one:
\begin{equation}
n_k=\frac{k_B T}{\epsilon_k-\mu}.
\eqname{class}
\end{equation}
Under this approximation, analytical results can be obtained for the
normal fraction $f_n$.

The mean density ${N}$ can be written as an infinite sum as:
\begin{equation}
{N}=\sum_k\frac{k_B T}{\frac{\hbar^2k^2}{2m}-\hbar k
  v-\mu}=\frac{mL^2k_B T}{2\pi^2 \hbar^2}\,\sum_{n\in{\mathbb Z}}
\frac{1}{n^2-2n{\tilde v}+\nu^2},
\eqname{CFN}
\end{equation}
where ${\tilde v}=v_{\rm rot}/v_1$ is the rescaled velocity
and $\nu^2=-m L^2 \mu/2\pi^2 \hbar^2$.
By applying the Poisson summation formula:
\begin{equation}
\sum_{n\in {\mathbb Z}}f(n)=\sum_{n\in{\mathbb Z}} {\hat f}(2\pi n),
\eqname{Poisson}
\end{equation}
where $f(x)$ is an arbitrary function and ${\hat  f}(k)=\int dx f(x)
\exp(-ikx)$ its Fourier transform, one is led to the final expression:
\begin{equation}
\eqname{CFN2}
{N}=\frac{m L^2 k_B T}{\hbar^2}\,\frac{1}{2 \pi \nu_0}\frac{\sinh(2\pi
  \nu_0)}{\cosh(2\pi\nu_0)-\cos(2\pi {\tilde v})},
\end{equation}
where $\nu_0$ is defined as $\nu_0^2=\nu^2-{\tilde v}^2$ \footnote{For
  $\nu^2<{\tilde v}^2$, which happens
when $L<[1-\cos(2\pi\tilde{v})]\rho\hbar^2/(m k_B T)$, $\nu_0$ is a
  purely imaginary quantity
and the analytical continuation of \eq{CFN2}  has to be taken.
}.
By applying the same Poisson summation formula \eq{Poisson} to the
mean momentum $\expect{P}$, one obtains:
\begin{equation}
\eqname{CFP}
\expect{P}=N m v-\frac{m L k_B T}{\hbar}\,\frac{\sin(2\pi{\tilde
    v})}{\cosh(2\pi \nu_0)-\cos(2\pi {\tilde v})}.
\end{equation}
Notice the periodicity of ${N}$ and $\expect{P}$ as functions
of the reduced velocity ${\tilde v}$. This in agreement with the
general result \eq{periodicP}.
 From \eq{CFN2} and \eq{CFP}  we immediately
obtain the generalized normal fraction:
\begin{equation}
\eqname{CFfn}
f_n=\frac{\expect{P}}{{N}m v}=1-\frac{\nu_0}{{\tilde
    v}}\,\frac{\sin(2\pi {\tilde v})}{\sinh(2\pi \nu_0)}.
\end{equation}

The physics of the non-interacting classical field is determined by the
  dimensionless parameter $\nu$ and the rotation velocity ${\tilde
  v}$. For a given rotation velocity $v_{\rm rot}$, the crossover from
  $\nu\gg 1$ to $\nu\ll 1$ corresponds to the Bose condensation in the finite
system.
In the high density $\nu\rightarrow 0$ limit where the gas is fully
  Bose-condensed, one has $\nu_0\rightarrow i{\tilde v}$ and the
  normal fraction \eq{CFfn} tends to zero. On the other hand, for
  $\nu\rightarrow \infty$, $\nu_0\rightarrow \infty$ and the normal
  fraction consequently tends to $1$.

In the absence of rotation ${\tilde v}=0$, Bose condensation in the finite system
  occurs for~\cite{atom_laser}:
\begin{equation}
k_B T\approx k_B T_{\rm BEC}=\frac{6 N \hbar^2}{m L^2},
\eqname{BEC}
\end{equation}
that is when the coherence length $l_c=\rho\lambda^2/2\pi$ of the gas
becomes larger than the size $L$ of the system.
Even in this zero-velocity case, the normal fraction
\begin{equation}
f_n^0=1-\frac{2\pi\nu_0}{\sinh(2\pi\nu_0)}
\eqname{CFfn0}
\end{equation}
differs from the non-condensed one $f_{\rm nc}$:
\begin{equation}
f_{\rm
  nc}=\frac{{N}-\expect{n_{k=0}}}{{N}}=1-\frac{\cosh(2\pi\nu_0)-1}
{\pi\nu_0\sinh(2\pi\nu_0)}.
\eqname{CFfnc}
\end{equation}
In particular, in the limit of a well-established Bose-Einstein
condensate $\nu_0\ll 1$, one has the peculiar relation:
\begin{equation}
\eqname{eqbizarre}
f_n^0=2f_{\rm nc},
\end{equation}
which states that the normal fraction is twice as large than the
non-condensed one, and, therefore, that the condensate is not
completely superfluid. This prediction is qualitatively opposite to
what is found in liquid $^4$He, which at low temperatures is totally
superfluid even if the condensate fraction is only of the order of
10\%~\cite{Helium1,Helium2}.

\subsection{Probability distribution of the total momentum}
For the ideal Bose gas, the total momentum is simply written in terms of the
single-particle occupation numbers $n_k$ as $P=\sum_k \hbar k\,n_k$,
so the characteristic function $g_p$ is equal to:
\begin{equation}
\eqname{g_P0}
g(\zeta)=\expect{e^{iz\sum_k\hbar k\,n_k}}=\prod_k\expect{e^{i\zeta\hbar
    k\,n_k}}=\prod_k
\frac{1-e^{-\beta(\epsilon_k-\mu)}}{1-e^{-\beta(\epsilon_k-\mu)}\,e^{i\zeta\hbar k}},
\end{equation}
which has a factorized form over single particle states.
In the classical field approximation, this infinite product can be calculated exactly,
as shown in section \ref{sec:ccm}.

\begin{figure}[htbp]
\begin{center}
\includegraphics[width=10cm,clip]{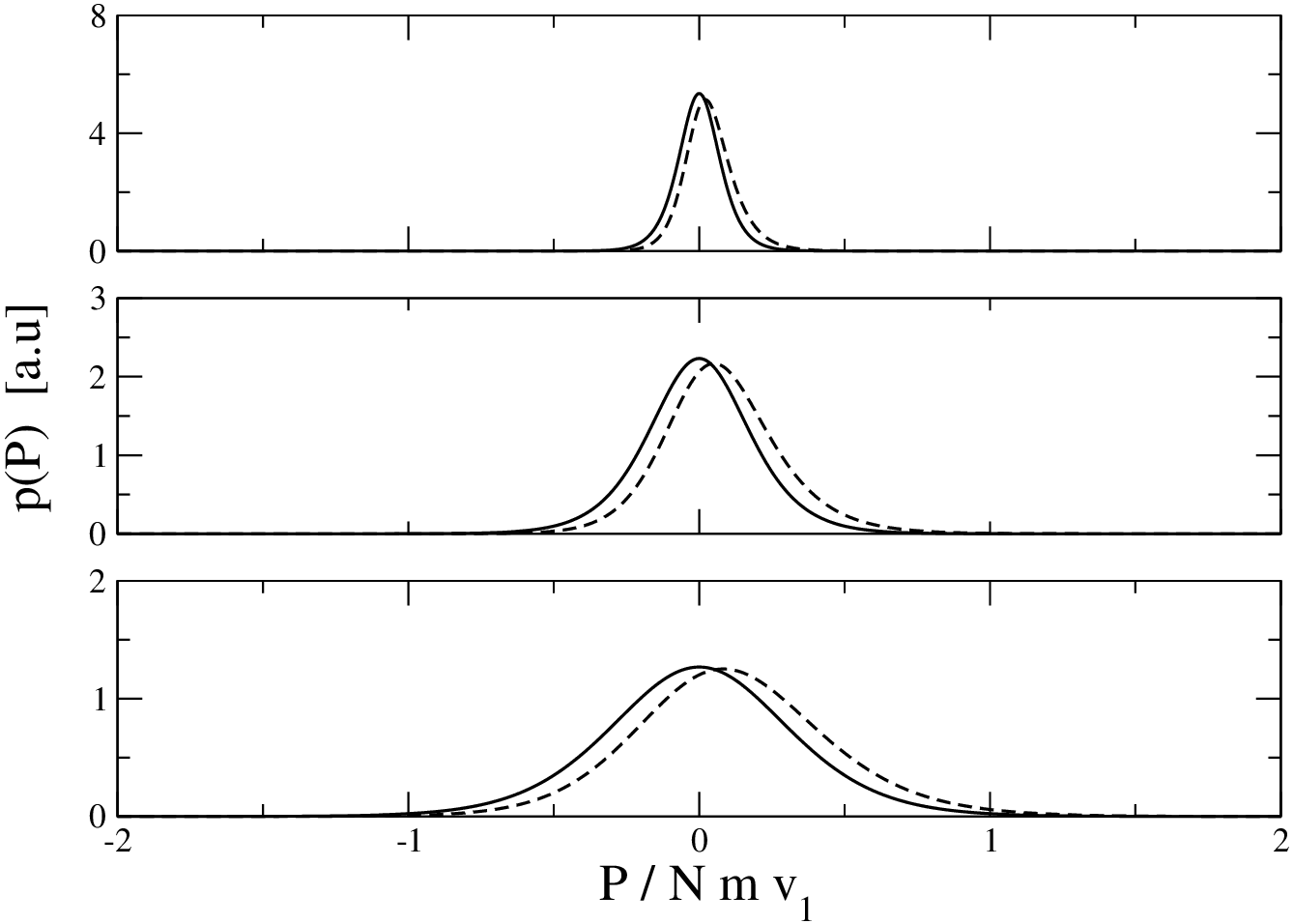}
\caption{Probability distribution $p(P)$ of the total momentum $P$ for
  an ideal Bose gas of $1000$ particles. Rotation velocity $v_{\rm
  rot}=0$ (solid lines), $v_{\rm rot}/v_1=0.125$ (dashed lines).
Temperatures $T/(N m v_1^2/2)=0.05, 0.125, 0.25$ (from top to bottom: the
  corresponding normal fractions are respectively: $f_n=0.3,
  0.64,0.92$).
\label{fig:distrP_NI}}
\end{center}
\end{figure}

The probability distribution $p(P)$ without the classical field approximation
is easily obtained by numerical Fourier transform
of \eq{g_P0} (fig.\ref{fig:distrP_NI}).
For a vanishing $v_{\rm rot}$ ,
the distribution $p(P)$ is simply
narrowed as the temperature is decreased and no
additional structure appears. An analytical proof of the fact that $p(P)$ has a single
maximum for $v_{\rm rot}=0$ is given in the appendix
\ref{appen:single_peak} in the classical field approximation.
Superfluidity effects can therefore be seen only
as a more rapid shrinking of the distribution as compared to the case
of a normal gas (cf. \eq{fn0}).
For a slow but finite rotation speed $v_{\rm rot}<v_1/2$, the flow of particles
simply appears as a shift-like distortion in the probability distribution
\footnote{Calculations are done here for the ideal Bose gas in the grand canonical
ensemble. In case where a condensate is present, this ensemble will give predictions
for $p(P)$ in agreement with the canonical ensemble only if the condensate
is in the mode $k=0$. Otherwise the non-physical grand canonical fluctuations
in the condensate mode will dramatically affect $p(P)$.}.

\section{Interacting gas I: multi-valley Bogoliubov approach}
\label{sec:BogoMV}

The present section is devoted to the calculation of the rotational
properties of a weakly interacting Bose gas at temperatures low enough
for density fluctuations to be weak. In this regime, the Bogoliubov
theory can be used to describe the system.
Since we are dealing with one-dimensional systems in which no
long-range order is present for the phase of the Bose field, the
extension of the Bogoliubov theory to quasi-condensates with only a
finite-range order should be used, as discussed in~\cite{Mora}.
For simplicity we shall use the version of the method in the canonical ensemble.

Quantitatively, the weakly interacting gas condition can be written in
terms of the density $\rho=N/L$ and the healing length $\xi$
as~\cite{WeaklyInter1D}:
\begin{equation}
\rho\xi=\sqrt{\frac{\hbar^2 \rho}{mg}}\gg 1,
\eqname{dilute}
\end{equation}
while the weakness of density fluctuations $\Delta\rho/\rho\ll 1$
requires~\cite{Mora}:
\begin{equation}
k_B T\ll k_B T_{\rm df}=\rho g \times \rho \xi=\Big(\frac{\hbar^2}{m}\rho^3 g\Big)^{1/2}.
\label{eq:Tdf}
\end{equation}
In the weakly interacting regime \eq{dilute}, $T_{\rm df}$ is always
much lower than the temperature 
\begin{equation}
T_{\rm deg}=2\pi \hbar^2 \rho^2/m
\eqname{Tdeg}
\end{equation}
for quantum degeneracy.

\subsection{Mean total momentum in the Bogoliubov approximation}
\label{subsec:moment_Bog}

The preliminary step to the Bogoliubov approach is to use the pure quasi-condensate
approximation and identify the quasi-condensate
wavefunction $\phi_0(x)$ as a local minimum of the Gross-Pitaevskii
energy functional:
\begin{equation}
E_{GP}[\phi]=\int\!dx\, \left[\frac{\hbar^2}{2m} |\partial_x \phi|^2
 +\frac{Ng}{2}|\phi|^4+i\hbar v_{\rm rot}\phi^*\partial_x \phi\right].
\eqname{GPEn}
\end{equation}
A local minimum has to be a stationary state of the Gross-Pitaevskii equation
\begin{equation}
\mu\phi = \left[-\frac{\hbar^2}{2m}\partial_x^2 + N g|\phi|^2 +i\hbar v_{\rm rot} \partial_x \right] \phi.
\end{equation}
Here we take for stationary states the usual form of plane waves:
\begin{equation}
\phi_0(x)=\frac{1}{\sqrt{L}}e^{ik_0x}.
\eqname{phi0}
\end{equation}
The periodic boundary conditions impose that the momentum
is quantized as $k_0=2\pi w_0/L$, $w_0$ being an integer called the
{\em winding number}. The corresponding velocity $v_0=w_0 v_1$ will be
called in the following {\em quasi-condensate} velocity.
The chemical potential is then
\begin{equation}
\mu=g\rho+\frac{1}{2}m\,v_0^2-m\,v_0\,v_{\rm rot}.
\end{equation}
The resulting energy of the pure quasi-condensate is then:
\begin{equation}
E_{GP}[v_{\rm rot},v_0]=N\Big[\frac{1}{2}m\,v_0^2-m\,v_0\,v_{\rm rot}\Big]+\frac{1}{2}N\rho g.
\eqname{GPEn0}
\end{equation}

Under the low-temperature and weak-interaction conditions stated
above, the fluctuations of the Bose field around the pure quasi-condensate
field can be described by the Bogoliubov
Hamiltonian~\cite{Mora,YvanHouches1}:
\begin{equation}
{\mathcal H}_{\rm Bog}=E_{GP}[v_{\rm rot},v_0]-\sum_{k\neq0} \epsilon_k\,
V_k^2+\sum_{k\neq0}\epsilon_k\bhd_k \bh_k,
\eqname{HBog}
\end{equation}
where the $\bh_k,\bhd_k$ operators are the usual (bosonic) destruction
and creation operators for the Bogoliubov quasi-particles. Their
dispersion relation $\epsilon_k(v_{\rm tot},v_0)$ is given by:
\begin{equation}
\epsilon_k(v_{\rm rot},v_0)=\epsilon^0_k+\hbar k (v_0-v_{\rm rot})
\eqname{epsilon}
\end{equation}
in terms of the usual dispersion relation $\epsilon^0_k$ for a system
at rest ($v_0=v_{\rm rot}=0$):
\begin{equation}
\epsilon^0_k=\Big[\frac{\hbar^2k^2}{2m}\big(\frac{\hbar^2k^2}{2m}+2\rho
  g\big)\Big]^{1/2}.
\eqname{epsilon0}
\end{equation}
The $V_k$ coefficients are defined as:
\begin{equation}
V_k=\frac{1}{2}\Big(\Big[\frac{\frac{\hbar^2 k^2}{2m}}{\frac{\hbar^2
      k^2}{2m}+2\rho g}\Big]^{1/4}-\Big[\frac{\frac{\hbar^2
      k^2}{2m}+2\rho g}{\frac{\hbar^2 k^2}{2m}}\Big]^{1/4}\Big).
\eqname{v_k}
\end{equation}
The fact that the $V_k$ do not depend on $v_0-v_{\rm rot}$ is a consequence of
Galilean invariance. One has also to calculate the total momentum operator (\ref{eq:Poper}):
\begin{equation}
P = N m v_0+\sum_{k\neq 0} \hbar k\, \hat{b}_k^\dagger \hat{b}_k.
\label{eq:Pjoli}
\end{equation}
The fact that the total number of particles enters in the first term of the righthand
side of this equation (rather than the number of particles in the quasi-condensate
or the number of particles in the superfluid fraction) is again a consequence of
Galilean invariance. A derivation of ${\mathcal H}_{\rm Bog}$ and $P$ is given
in the appendix \ref{appen:Bog}.

The quasi-condensate mode $\phi_0(x)$ is a local minimum of the Gross-Pitaevskii
energy functional if all the $\epsilon_k$ are positive, which guarantees
the thermodynamical stability of the gas. This
condition can be reformulated as:
\begin{equation}
|v_0-v_{\rm  rot}|\leq\textrm{min}_{k\neq 0}\Big[\frac{\epsilon^0_k}{\hbar k}\Big].
\eqname{Landau}
\end{equation}
When the length $L$ is larger than the healing length $\xi$, this criterion reduces to the usual Landau
criterion which states that the flow velocity as measured in
the moving frame has to be smaller than the sound velocity $c=(\rho g/m)^{1/2}$ of
the gas at rest.

A key assumption of our multi-valley Bogoliubov approach is that the
density matrix of the system is a statistical mixture of states of different
winding numbers $w_0$ and thus quasi-condensate velocities $v_0=w_0
v_1$. Each value of $v_0$ corresponds to a separate
minimum -- a {\em valley} -- of the Gross-Pitaevskii energy functional
\eq{GPEn} and the fluctuations around each of them are treated within
the Bogoliubov approximation previously discussed.

The expectation value of any observable, e.g. the momentum $P$, is
then expressed as an average over the different valleys:
\begin{equation}
\expect{P}=\frac{1}{{\mathcal Z}_{T}} \sum_{v_0}{\mathcal
  Z}_{v_0}\,\expect{P}_{v_0}.
\eqname{ExpectP}
\end{equation}
Obviously, only stable states satisfying \eq{Landau} are to be taken
into account. The ${\mathcal Z}_{v_0}$ are the contributions of the
different valleys to the total partition function:
\begin{equation}
{\mathcal Z}_{v_0}=\textrm{Tr}[e^{-\beta {\mathcal H}}]_{v_0}=A\;
e^{-\frac{1}{2}\beta N m(v_0-v_{\rm rot})^2}\prod_{k\neq
  0}\frac{1}{1-e^{-\beta \epsilon_k(v_{\rm rot},v_0)}},
\eqname{Z_w_0}
\end{equation}
where $A$ is a overall factor which does not depend on $w_0$ and
therefore drops out from all the calculations that follow.
The total partition function is ${\mathcal Z}_T=\sum_{v_0} {\mathcal
  Z}_{v_0}$.

The expectation value of $P$ within a given valley $\langle P\rangle_{v_0}$ is obtained
as the thermal average of (\ref{eq:Pjoli}):
\begin{equation}
\langle P\rangle_{v_0} = N m v_0 +
\sum_{k\neq 0} \frac{\hbar k}{\exp\{-\beta[\epsilon_k^0-\hbar k (v_{\rm rot}-v_0)]\}-1}
\equiv  N m v_0 + Nm (v_{\rm rot}-v_0) \; f_n^v (v_{\rm rot}-v_0).
\label{eq:P2}
\end{equation}
The function $f_n^v(v)$ has a very simple physical interpretation: it is
the generalized normal fraction of the gas that one would predict {\sl in a single valley Bogoliubov treatment}
(i.e. winding number $w_0=0$) for a rotation velocity $v$. Hence the
superscript $v$.

In the remaining part of this section we shall restrict to the case
\begin{equation}
L\gg \xi.
\label{eq:L_vs_xi}
\end{equation}
This will provide a considerable simplification of the expression
of $\langle P \rangle$.

First because this allows to replace in $f_n^v(v)$ the sum over $k$ by
an integral. The resulting integral may be calculated in the limit
of low and high temperatures.
If the temperature is sufficiently low:
\begin{equation}
k_B T\ll \rho g \Big(1-\Big|\frac{v}{c}\Big|\;\Big),
\eqname{LowT}
\end{equation}
only the linear part of the spectrum effectively contributes to the
integral.  This leads to
\begin{equation}
f_n^v(v)=\frac{(k_B T)^2}{c^3}\frac{\pi}{3m\hbar \rho}\frac{1}{(1-{v^2}/{c^2})^2},
\eqname{f_nLowT}
\end{equation}
where $c=\sqrt{{\rho g}/{m}}$ is the sound velocity.

For temperatures $k_B T\gg \rho g$ but still smaller than the
temperature $T_{\rm df}$ at which density
fluctuations becomes important, a
classical field approximation can be performed on the Bose
distribution law, which leads to
\begin{equation}
f_n^v(v)=\frac{k_B T}{\hbar c \rho} \frac{1}{\sqrt{1-{v^2}/{c^2}}}.
\eqname{f_nHighT}
\end{equation}
Within their own domain of validity, both \eq{f_nLowT} and
\eq{f_nHighT} predict a single-valley normal fraction much smaller than one.

Let us assume that $|v_{\rm rot}-v_0|\ll c$ for all the relevant terms in the sum
(\ref{eq:ExpectP}).
In this case, \eq{P2} can be simplified as follows:
 \begin{equation}
\langle P\rangle_{v_0}\simeq N m f_n^{v,0} v_{\rm rot}+N m v_0 (1-f_n^{v,0}),
\eqname{P3}
\end{equation}
where $f_n^{v,0}$ is the zero-rotation limit of $f_n^v$:
\begin{equation}
f_n^{v,0} = f_n^v(0) =
\frac{1}{Nm k_B T}\sum_{k\neq 0}(\hbar k)^2 \frac{e^{\beta\epsilon_k^0}}{(e^{\beta\epsilon_k^0}-1)^2}.
\eqname{f_nBog2}
\end{equation}
The result \eq{P3} can be physically understood in the following
terms: for any value of the quasi-condensate velocity $v_0$, the normal fraction $f_n^{v,0}$ within the
valley is moving at a rotation velocity $v_{\rm rot}$, while the
superfluid fraction $1-f_n^{v,0}$ moves independently at $v_0$.

In this limit $|v_{\rm rot}-v_0|\ll c$, the expression \eq{Z_w_0} for the partial partition
function for each valley can be simplified by expanding the product
to second order in $v_{\rm rot}-v_0$:
\begin{equation}
\eqname{product_k}
\prod_{k\neq 0}\frac{1}{1-e^{-\beta
  \epsilon_k(v_{\rm rot},v_0)}}
\simeq e^{\frac{1}{2}\beta N m
  f_n^{v,0}(v_{\rm rot}-v_0)^2} \prod_{k\neq
  0}\frac{1}{1-e^{-\beta \epsilon^0_k}}.
\end{equation}
By inserting this expression into \eq{Z_w_0}, the final expression for the
expectation value of the momentum \eq{ExpectP} gets the simple form:
\begin{equation}
P(v_{\rm rot})=f_n^{v,0} N m v_{\rm rot}+N (1-f_n^{v,0})\,\sum_{v_0} q(v_0)\, m v_0,
\eqname{ExpectPBogMV}
\end{equation}
in which the normal gas always moves at a velocity $v_{\rm rot}$ and the
probability distribution $q(v_0)$ for the quasi-condensate to have a
velocity $v_0$ has the Gaussian form:
\begin{equation}
q(v_0)={\mathcal N}_q\,\exp\Big[-\frac{m N (1-f_n^{v,0}) }{2k_B T} \big(v_{\rm rot}-v_0\big)^2\Big],
\eqname{Probv_0}
\end{equation}
where ${\mathcal N}_q$ is the normalization factor.
Note that only the fraction $1-f_n^{v,0}$ of atoms
which are superfluid in a single valley treatment is actually involved
in \eq{Probv_0}.
The normal fraction of the gas can be obtained
from (\ref{eq:ExpectPBogMV}):
\begin{equation}
f_n^{0} = f_n^{v,0} + \frac{N m (1-f_n^{v,0}) ^2}{k_B T}
\frac{\sum_{v_0} v_0^2 \exp\Big[-\frac{1}{2}\beta m N (1-f_n^{v,0})\, v_0^2 \Big]}
{\sum_{v_0} \exp\Big[-\frac{1}{2}\beta m N (1-f_n^{v,0})\, v_0^2 \Big]}.
\end{equation}
Remarkably this expression coincides with formula (11) of \cite{Svistunov}
if one takes $\rho_S=(1-f_n^{v,0})\rho$ in \cite{Svistunov}.

We briefly come back to the condition $|v_{\rm rot}-v_0|\ll c$.
For $v_0=0$, this hypothesis is a direct consequence of
(\ref{eq:L_vs_xi}) since $v_{\rm rot}$ can be taken in the interval $[-v_1/2,v_1/2]$.
What happens if $v_0 \gg v_1$?
The maximal accessible value $v_0^{\rm max}$ of $v_0$ allowed by the probability
distribution \eq{Probv_0} corresponds to $mv_0^2/2\approx k_B
T/N$. $v_0^{\rm max}$ is much smaller than $c$ provided $k_B T\ll N
\rho g$. As:
\begin{equation}
\frac{k_B T}{N \rho g}=\frac{T}{T_{\rm df}}\frac{\xi}{L},
\end{equation}
this is automatically satisfied since we have assumed $L\gg \xi$ and
$T\ll T_{\rm df}$.

Even if the normal fraction $f_n^{v,0}$ within a single valley is always small
$f_n^{v,0}\ll 1$ in the regime of validity of the Bogoliubov approximation,
the true normal fraction tends to one as soon as
several valleys are thermally populated
\begin{equation}
k_B T \geq k_B T_v=\frac{1}{2} N m v_1^2.
\label{eq:tv}
\end{equation}
In this case, one can replace the discrete sum over $v_0$ by an integral
in  (\ref{eq:ExpectPBogMV}):
\begin{equation}
\langle P\rangle \simeq N m v_{\rm rot},
\label{eq:P_T_l_Tv}
\end{equation}
finding that the transport of atoms in the presence of a finite
$v_{\rm rot}$ is essentially due to the redistribution of the
population among the different valleys.
In other words, as several different values of the quasi-condensate
velocity $v_0$ are accessible, the presence of a finite rotation
velocity causes an imbalance of the relative probabilities $q(v_0)$
and therefore a net matter flow.

\begin{figure}[htbp]
\begin{center}
\includegraphics[width=10cm,clip]{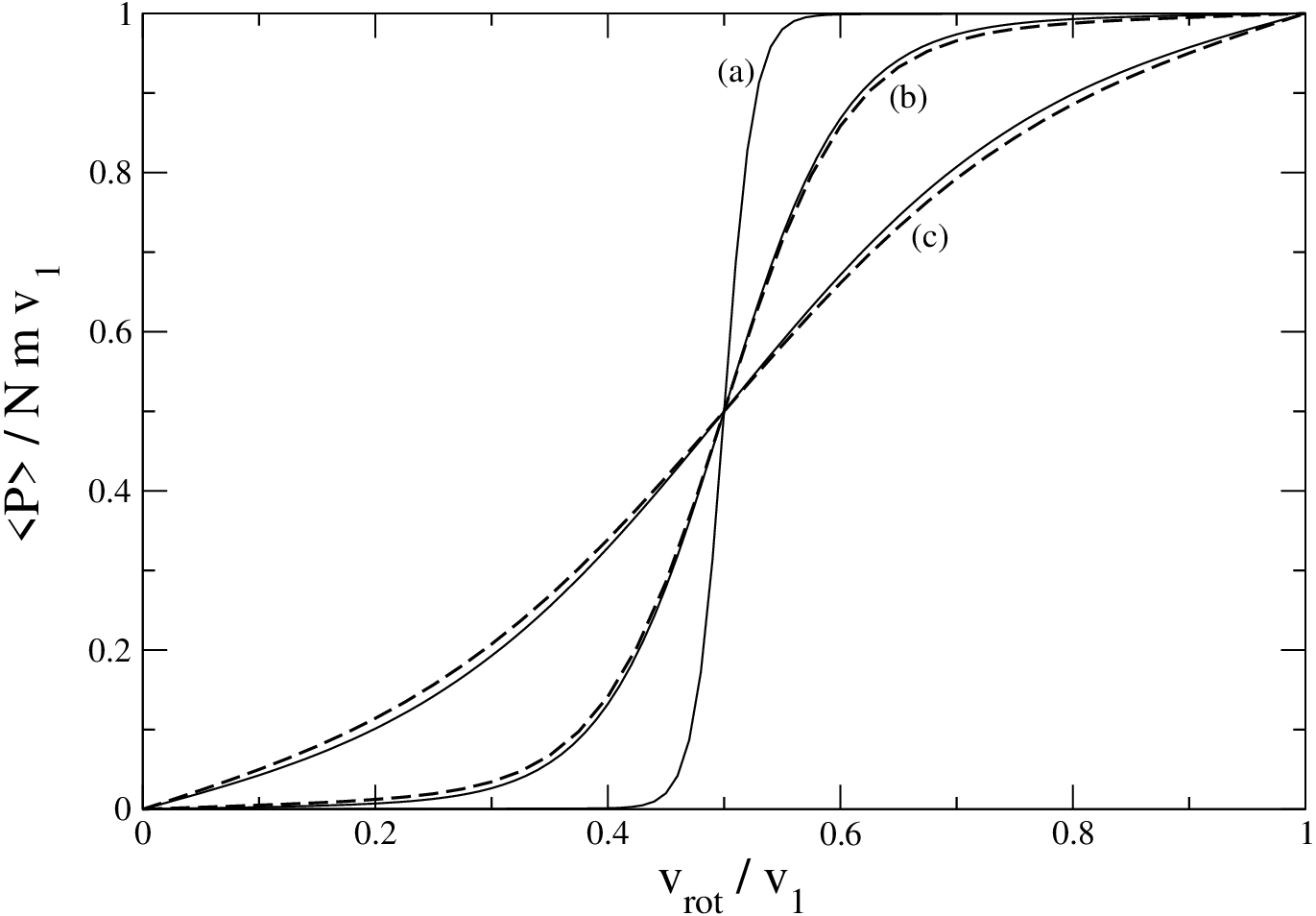}
\caption{Average value of the total momentum $P$ as a function of the
  rotation velocity $v_{\rm rot}$ for a Bose gas of $1000$ interacting
  atoms. Chemical potential $\rho g/k_B T_v =0.1$. Temperatures: $T/T_v=0.025$ (a), $0.1$ (b),
  $0.25$ (c). Solid lines:
  many-valley Bogoliubov predictions. 
Dashed lines for (b) and (c): full numerical solution of the classical field
model
of  section \ref{sec:ccm}.  For curve (a), $k_B T = \rho g /4 \ll \rho
  g$, outside the applicability range of the classical field model.
\label{fig:ptot}}
\end{center}
\end{figure}

The Bogoliubov prediction for $\langle P\rangle$ as function of $v_{\rm rot}$
is plotted in figure~\ref{fig:ptot} for various values of the temperature.
In the low temperature regime $T \ll T_v$, only the valley
with $w_0=0$ is generally occupied for $v_{\rm rot} < v_1/2$:
the full normal fraction is close to the single
valley prediction, so that the gas is almost fully superfluid.
When $v_{\rm rot}$ approaches $v_1/2$,
the valleys $w_0=0$ and $w_0=1$ become
nearly degenerate: the
Gross-Pitaevskii energies \eq{GPEn0} of a condensate at rest $v_0=0$ and in the first
excited state $v_0=v_1$ are nearly equal.
In this case, even if $T \ll T_v $, both valleys can be
populated and their relative weights will be strongly
dependent on $v_{\rm rot}$, see the step function in
figure~\ref{fig:ptot}. For $v_{\rm rot}>v_1/2$, only the valley
$w_0=1$ is occupied.
The width of the crossover from a value $\langle P\rangle \ll N m
v_{\rm rot}$ to a value $N m v_{\rm rot}-\langle P\rangle \ll N m
(v_1-v_{\rm rot})$ is of the order of
\begin{equation}
\eqname{crossover}
\Delta v_{\rm rot}=\frac{k_B T}{N m v_1}.
\end{equation}
In the opposite regime of a temperature on the order of $T_v$, where several
valleys are thermally populated, the step behaviour of
$P(v_{\rm rot})$ is smoothed out and the gas become normal, as predicted in
(\ref{eq:P_T_l_Tv}).

\subsection{Comparison with Quantum Monte Carlo simulation}
\label{sec:QMC}

The predictions of the many-valley Bogoliubov approach can be verified
against a Quantum Monte Carlo calculation
performed using a recently developed stochastic field technique for
the interacting Bose gas~\cite{GPstochT}.
As rotation is described by an additional single-particle term
$-Pv_{\rm rot}$ in the Hamiltonian, the numerical simulation does not
present any further complication with respect to the ones previously
performed, e.g. for the study of the statistical properties of a
Bose-Einstein condensate~\cite{StatN0}.

\begin{figure}[htbp]
\begin{center}
\includegraphics[width=10cm,clip]{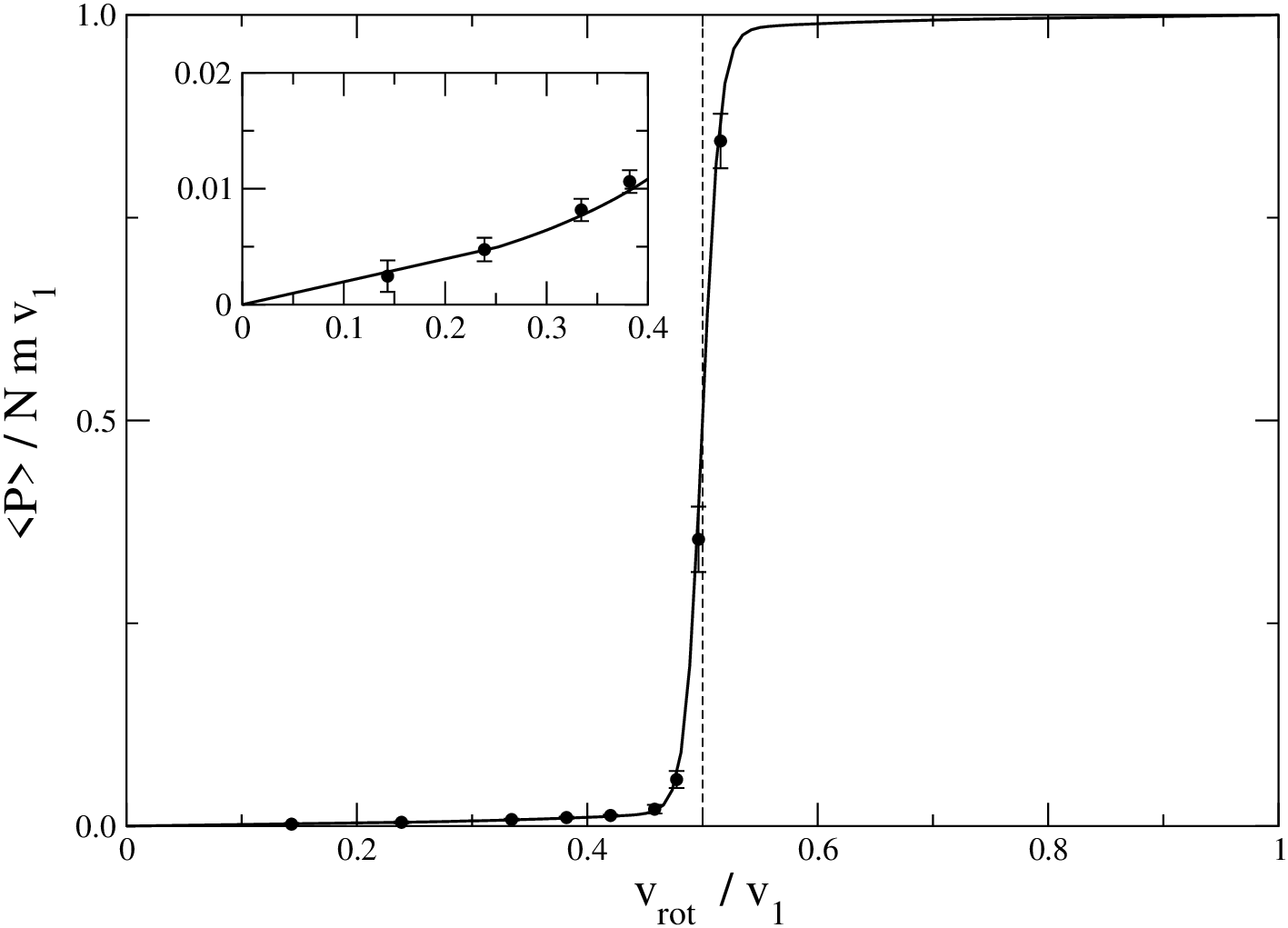}
\caption{Average value of the total momentum $P$ as a function of the
  rotation velocity $v_{\rm rot}$ for a 1D Bose gas of $42$ interacting
  atoms. Chemical potential $\rho g/k_B T_v=0.017$, $T/T_v=0.017$.
Solid line: many-valley Bogoliubov theory.
Points: Quantum Monte Carlo calculation.
\label{fig:QMC}}
\end{center}
\end{figure}

In fig.\ref{fig:QMC} we have compared the result of the Monte Carlo
simulation with the prediction of the many-valley
Bogoliubov theory. As the condition $L\gg\xi$ is not fulfilled, a
numerical evaluation of (\ref{eq:ExpectP}-\ref{eq:P2}) is
necessary.
The parameters of the figure correspond to a regime in which only a
single valley is thermally populated $T\ll T_v$: the agreement is good
in the slow-velocity regime
$v_{\rm rot}\ll v_{1}$ as well as close to the point $v_{\rm rot}=v_{1}/2$
where the two valleys of winding numbers respectively $w_0=0$ and $w_0=1$ become
degenerate and the system is in a statistical mixture of them.

\subsection{Probability distribution of the total momentum}

If the temperature is sufficiently low for density
fluctuations to be small, the many-valley Bogoliubov theory introduced
in the previous subsection can be used to calculate not only the
normal fraction but also the complete probability distribution $p(P)$
for the total momentum $P$.

Using the expression (\ref{eq:Pjoli}) of the total momentum operator
in the Bogoliubov approximation,
the characteristic function $g(\zeta)$ can be written as:
\begin{equation}
\eqname{g_PBog}
g(\zeta)=\frac{1}{{\mathcal Z}_T}\sum_{v_0}{\mathcal Z}_{v_0} e^{i\zeta
  N m v_0} \expect{e^{i\zeta\sum_k\hbar k
    n_k}}_{v_0}=
\frac{1}{{\mathcal Z}_T}\sum_{v_0} e^{-\frac{1}{2} \beta N m (v_0-v_{\rm
    rot})^2}
e^{i\zeta N    m v_0}
\prod_k
\frac{1}{1-e^{-\beta
    [\epsilon_k^0-\hbar k (v_{\rm rot}-v_0)]}e^{i\zeta\hbar k}},
\end{equation}
where $n_k$ are now the occupation numbers of the Bogoliubov
quasi-particle modes of energies $\epsilon_k$.

It is apparent in the last factor of (\ref{eq:g_PBog}) that
the presence of $\zeta$ can be reinterpreted as a complex
shift of $v_{\rm rot}-v_0$ by an amount $-i k_B T \zeta$.
If we assume that
\begin{equation}
|v_{\rm rot}-v_0-i k_B T \zeta| \ll c,
\label{eq:hypo_zeta}
\end{equation}
we can simplify \eq{g_PBog} in exactly the same way
which led to \eq{product_k}:
\begin{equation}
\eqname{g_PBog2}
g(\zeta)\simeq\frac{1}{{\tilde {\mathcal Z}}_T}\sum_{v_0}e^{-{\beta N
    m}(1-f_n^{v,0})(v_{\rm rot}-v_0)^2/2} e^{i\zeta m N
  v_0(1-f_n^{v,0})}
e^{i\zeta m N v_{\rm rot} f_n^{v,0}} e^{-{mN f_n^{v,0} k_B T}\zeta^2/2}.
\end{equation}
The distribution function $p(P)$ for the momentum can then be calculated by
Fourier transform of the characteristic function \eq{g_PBog2}:
\begin{equation}
\eqname{p_PBog}
p(P)=\frac{1}{\Pi_T}\sum_{v_0} e^{-{\beta m N
    (1-f_n^{v,0})}(v_0-v_{\rm rot})^2/2}
e^{- [P- m v_0 N (1-f_n^{v,0})-m v_{\rm rot} N f_n^{v,0}]^2/
(2m N f_n^{v,0} k_B T)}.
\end{equation}
The structure of \eq{p_PBog} is physically transparent and can be
summarized as follows:
\begin{itemize}

\item Each valley contributes as a peak of width $(f_n^{v,0}N m k_B T)^{1/2}$
in accord with \eq{fn0} as applied to a single valley.

\item The peak corresponding to each valley is centered at the value $m v_0
N (1-f_n^{v,0})+m v_{\rm rot} f_n^{v,0}$ of the momentum: as expected,
the (single-valley) superfluid fraction $(1-f_n^{v,0})$ moves at the
quasi-condensate velocity $v_0$, while the normal one $f_n^{v,0}$ moves
at $v_{\rm   rot}$.

\item The occupation probability of each valley is proportional to a
  Gaussian involving the kinetic energy in the rotating frame of the
(single-valley) superfluid fraction.
\end{itemize}

So, for temperatures such that:
\begin{equation}
T_v \ll T \ll \frac{T_v}{f_n^{v,0}}
\label{eq:regime}
\end{equation}
the probability distribution  of momentum $p(P)$ is given by a series of
narrow isolated peaks.

\begin{figure}[htpb]
\begin{center}
\includegraphics[width=10cm,clip]{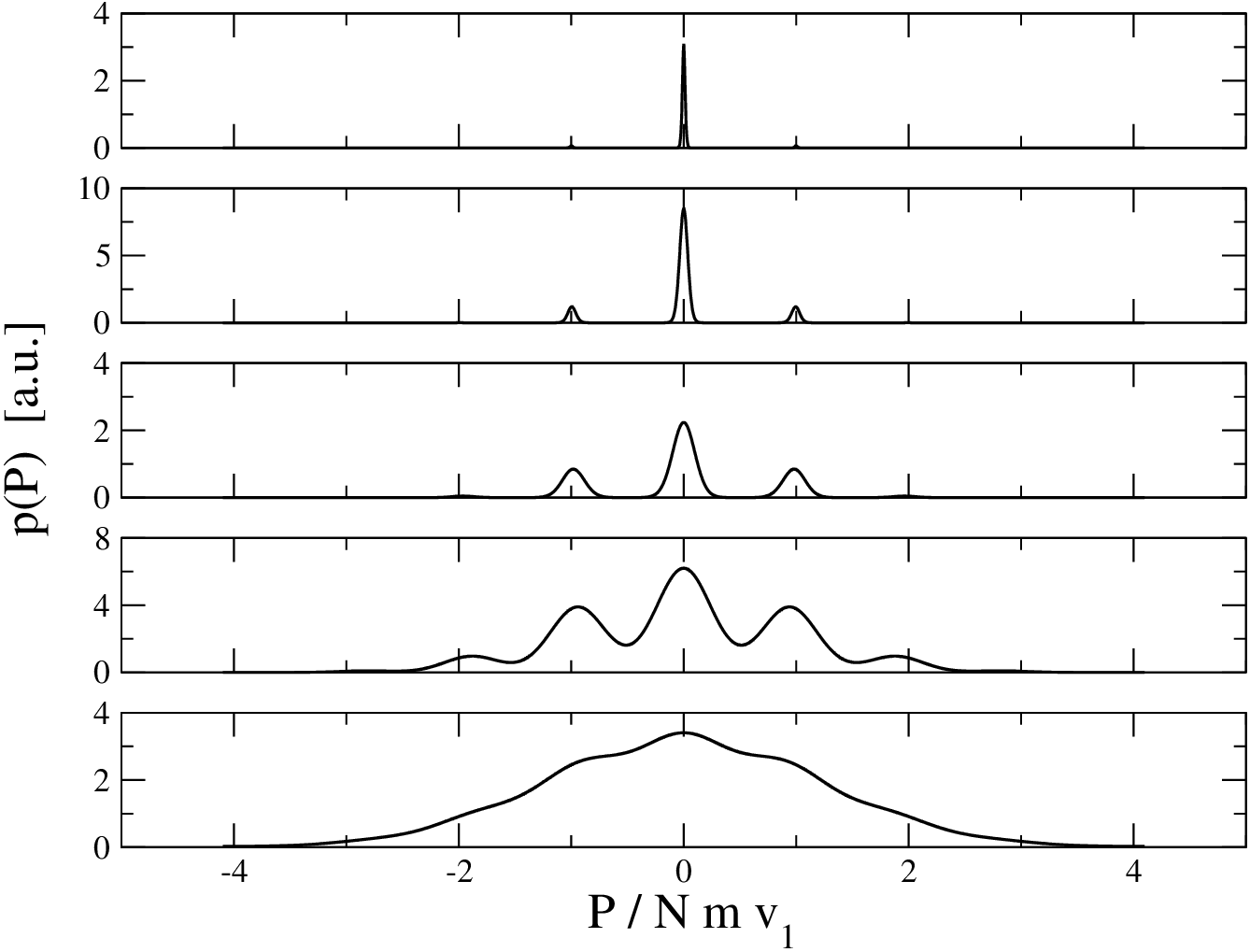}
\caption{Multi-valley Bogoliubov prediction for the probability
  distribution $p(P)$ of the total momentum $P$ for
  an interacting Bose gas of $1000$ particles. Rotation velocity $v_{\rm
  rot}=0$. Chemical potential $\rho g/k_B T_v=3.2$.
Temperature: $T/T_v=0.25,0.5,1,2,3$ (from top to bottom).
\label{fig:distrP_I}}
\end{center}
\end{figure}

The probability distribution $p(P)$ is plotted in
fig.\ref{fig:distrP_I} for $v_{\rm rot}=0$ and different values of the
temperature.
At low temperatures $T\ll T_v$, only a single valley is
populated and $p(P)$ shows a single narrow peak around $P=0$. In this
case, the gas is nearly fully superfluid.
For temperatures growing across $T/T_v \approx 1$, higher valleys start to be
populated as well, more peaks become visible and the width of each of
them gets larger. Correspondingly, the superfluid fraction decreases
to zero.
At sufficiently high temperature, the isolated peaks
merge into a broad, unstructured, distribution.
Note that for all
temperatures in the figure, the single-valley normal fraction
$f_n^{v,0}$ remains always significantly smaller than unity.

\begin{figure}[htbp]
\begin{center}
\includegraphics[width=10cm,clip]{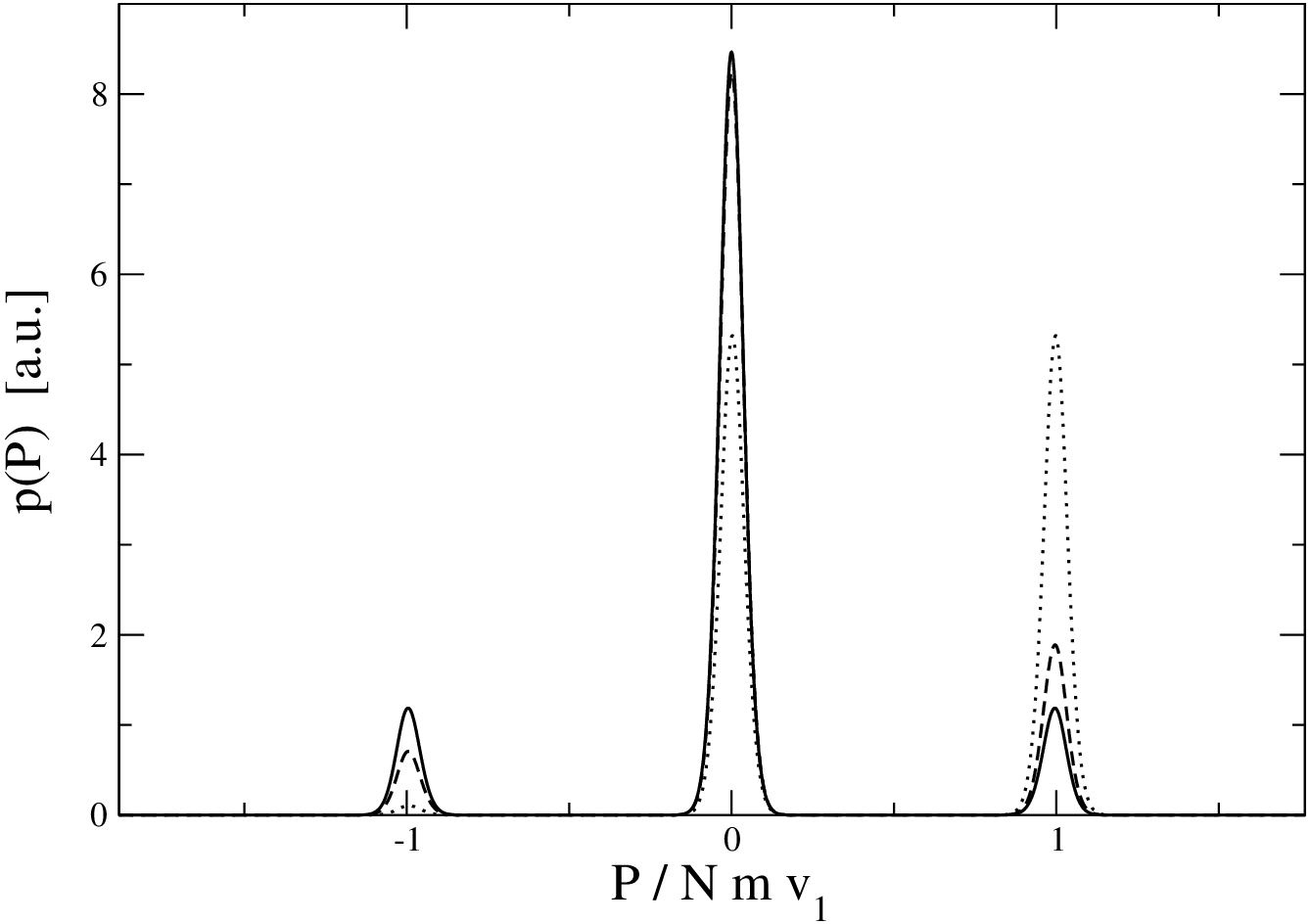}
\caption{
Multi-valley Bogoliubov prediction for the probability distribution
$p(P)$ of the total momentum $P$ for
  an interacting Bose gas of $1000$ particles. Rotation velocity $v_{\rm
  rot}/v_1=0$ (solid), $0.125$ (dashed), $0.5$ (dotted). Chemical
  potential $\rho g/k_B T_v=3.2$.
Temperature $T/T_v=0.5$.
\label{fig:distrP_Iv}}
\end{center}
\end{figure}

Conventional superfluidity occurs only when the width of the envelope becomes
of the order of the spacing between peaks, so that only the one whose
quasi-condensate velocity $v_0$ is closest to the rotation velocity
$v_{\rm rot}$ results effectively populated.
Otherwise, mass transport occurs in the presence of a finite rotation
velocity $v_{\rm rot}$ as a consequence of the redistribution of
population between the different valleys more than of the distortion
within a given valley.
This behaviour can be clearly observed in fig.\ref{fig:distrP_Iv},
where the effect of a finite $v_{\rm rot}$ on the probability
distribution $p(P)$ is shown: although the shape of each single peak
is weakly affected, the envelope function suffers a dramatic distortion.

Let us briefly discuss the validity condition of (\ref{eq:hypo_zeta}).
The validity condition of $|v_{\rm rot}-v_0|\ll c$ was already discussed
in subsection \ref{subsec:moment_Bog}. What is left is to compare $k_B T |\zeta|$ to $c$.
The narrowest features in $p(P)$ are found to be the peaks corresponding to
single valley normal fractions, of momentum width  $(f_n^{v,0}Nmk_B T)^{1/2}$.
As $g(\zeta)$ is the Fourier transform of $p(P)$, this results in a
maximal value of $\zeta$ on the order of
\begin{equation}
\zeta_{\rm max} = \frac{1}{(f_n^{v,0}Nmk_B T)^{1/2}}.
\end{equation}
In the regime $k_B T \gg \rho g$, $f_n^{v,0}$ is given by the $v=0$
limit of \eq{f_nHighT}. We then obtain
\begin{equation}
\frac{k_B T \zeta_{\rm max}}{c} \simeq \frac{1}{(L/\xi)^{1/2}} \ll 1
\end{equation}
since we assumed $L\gg \xi$.
In the regime $k_B T \ll \rho g$, we use \eq{f_nLowT} with $v=0$ to obtain
\begin{equation}
\frac{k_B T \zeta_{\rm max}}{c}\simeq \left(\frac{\hbar c}{L k_B T}\right)^{1/2}.
\end{equation}
The condition (\ref{eq:hypo_zeta}) then imposes a lower bound on the temperature,
\begin{equation}
k_B T \gg \hbar c/L.
\end{equation}
Since $\hbar c/L$ is the energy of the lowest Bogoliubov excitation in
a single valley, this condition is equivalent to a non-zero temperature
regime in the valley. Note that this condition is automatically satisfied
in the multi-valley regime $T \geq T_v$ for a weakly
interacting Bose gas.

\section{Interacting gas II: classical field model}
\label{sec:ccm}

The results of the previous section have been obtained on the grounds
of a Bogoliubov theory which assumes that the density fluctuations are
weak; this condition requires the temperature to be sufficiently low
$T\ll T_{\rm df}$ where $T_{\rm df}$ is given in (\ref{eq:Tdf}).
On the other hand, for temperatures larger than the chemical potential
but still lower than the degeneracy temperature
\begin{equation}
k_B T_{\rm deg} \gg k_B T\gg \rho g,
\eqname{TforCFT}
\end{equation}
a classical field model can be applied. In the present section, we
shall discuss the predictions of this approach regarding the
superfluidity properties of the gas.
The actual existence of a temperature range \eq{TforCFT} is guaranteed
by the weak interaction condition \eq{dilute}. In this regime, as one can
see in the diagram in fig.\ref{fig:T_diagram}, the applicability
domains of classical field and Bogoliubov theories have a
non-vanishing overlap and, in particular, give coincident predictions,
as we shall see.

\subsection{The model and its solution}
We generalize to the rotating case
the classical field model of \cite{atom_laser,Scalapino,houches03}. In this generalization
of the model,
the complex field $\psi(z)$ has a grand canonical thermal equilibrium distribution
$P[\psi]$ proportional
to $\exp(-\beta E[\psi])$ where $E[\psi]$ is the Gross-Pitaevskii energy
functional:
\begin{equation}
E[\psi] = \int_0^{L} dz\, \left[\frac{\hbar^2}{2m} |\partial_z \psi|^2 +\frac{g}{2}|\psi|^4
-\frac{\hbar}{i} v_{\rm rot} \psi^*\partial_z\psi -\mu |\psi|^2\right]
\end{equation}
restricting to the configurations of the complex field obeying the boundary
condition $\psi(0)=\psi(L)$.

Expectation values of quantum observables are
obtained by replacing $\hat{\psi}$ with $\psi$, $\hat{\psi}^\dagger$ with $\psi^*$
and then averaging over the thermal distribution $P[\psi]$. At this stage, the reader may
argue that a classical field thermal distribution is expected to lead to divergences
in the observables in the absence of an energy cut-off, reminiscent of the black-body
catastrophe of 19th century. A very fortunate consequence
of the 1D character of the gas is that the classical field model gives
finite predictions for the observables relevant for this paper, like the
mean density, the mean momentum of the gas, the probability distribution of
the total momentum of the gas, as we shall see. This suppresses the issue
of an energy cut-off dependence.

Calculation of expectation values can be performed exactly in the classical field model:
the summation over all possible complex paths $z\rightarrow \psi(z)$ can be viewed
formally as a Feynman path integral over trajectories of a single quantum particle
in 2D, $z$ playing the role of a fictitious time, the real and imaginary parts
of $\psi$ corresponding to fictitious coordinates $x$ and $y$. Using in the reverse order
the Feynman formulation of quantum mechanics, one can map the functional integral
over all paths into a Feynman propagator for a fictitious Hamiltonian
of a quantum particle moving in 2D, here in {\sl imaginary} time.
More details are given in \cite{atom_laser,houches03}. We give here
without proof the expression of the fictitious Hamiltonian for a
rotating system: 
\begin{equation}
{\mathcal H}_{v_{\rm rot}} = \frac{p_x^2+p_y^2}{2M} +i \frac{m {v_{\rm rot}}}{\hbar} L_z
+\frac{1}{2}\hbar \beta g (x^2+y^2)^2 -\hbar \beta
(\mu+\frac{1}{2}m{v_{\rm rot}}^2)
(x^2+y^2)
\label{eq:hamil_clas}
\end{equation}
where the fictitious mass is
\begin{equation}
M= \frac{\hbar^3}{mk_B T},
\end{equation}
$p_x,p_y$ are the momentum operators of the fictitious particle along $x,y$
and $L_z = x p_y-y p_x$ is the angular momentum operator of the particle along
$z$. Note that the fictitious Hamiltonian is not hermitian for ${v_{\rm rot}}\neq 0$,
but its anti-hermitian part commutes with its hermitian part,
which is indeed rotationally invariant. A numerical diagonalization of the hermitian
part of ${\mathcal H}$
is therefore very simple, as one has to solve a Schr\"odinger equation for the
radial part of the eigenfunctions only. The corresponding eigenvectors are labeled
by two quantum numbers, the angular momentum $l\in \mathbb{Z}$ and the
radial quantum number $n\in\mathbb{N}$,
and $E_{n,l}$ is the corresponding real eigenvalue.

\subsection{Exact expressions for relevant observables}
We give the explicit expression for some useful expectation values.
The calculation of the mean density in the classical field model
is required to determine the chemical potential $\mu$ for a
given mean total number of particles. The mean density is given by
\begin{equation}
\langle \psi^*\psi \rangle = \langle x^2+y^2\rangle_q
\label{eq:dens_clas}
\end{equation}
where the ``quantum expectation value" of any operator $O$
for the fictitious particle is defined as
\begin{equation}
\langle O \rangle_q \equiv
\frac{\mbox{Tr}\left[O\,e^{-L{\mathcal H}_{v_{\rm rot}}/\hbar}\right]}{\mbox{Tr}\left[e^{-L{\mathcal H}_{v_{\rm rot}}/\hbar}\right]}.
\end{equation}
The characteristic function for the total
momentum $P=-i\hbar\int \psi^*\partial_z\psi$ is
\begin{equation}
g(\zeta) =  \langle e^{i\zeta P}\rangle =
\frac{\mbox{Tr}\left[e^{-L{\mathcal H}_{w}/\hbar}\right]}{\mbox{Tr}\left[e^{-L{\mathcal H}_{{v_{\rm rot}}}/\hbar}\right]}
\label{eq:gen_clas}
\end{equation}
where we have introduced the complex velocity
\begin{equation}
w={v_{\rm rot}}-i\zeta k_B T
\eqname{w}
\end{equation}
and ${\mathcal H}_{w}$ is obtained by replacing $v_{\rm rot}$ with $w$ in ${\mathcal H}_{v_{\rm rot}}$.
 The mean total momentum of the gas is related to the derivative of the characteristic function
in $\zeta=0$:
\begin{equation}
\langle P\rangle = -i \frac{dg}{d\zeta}(0) = N m {v_{\rm rot}} - i\frac{m k_B T}{\hbar^2} L \langle L_z\rangle_q
\label{eq:P_clas}
\end{equation}
where we used (\ref{eq:dens_clas}) to obtain the mean total number of particles $N$.
The first order expansion in $v_{\rm rot}$ of this expression, when combined with
the $v_{\rm rot}\rightarrow 0$ limit of (\ref{eq:fn}), leads to the exact expression
for the standard definition of the normal fraction \cite{chemi}:
\begin{equation}
f_n^0 = 1 - \frac{m k_B T L}{\rho \hbar^2} \frac{\langle L_z^2\rangle_q(v_{\rm rot}=0)}{\hbar^2}.
\label{eq:fn_class}
\end{equation}
Note that this expression makes explicit the fact that one has always $f_n^0 \leq 1$, which
justifies the name of normal `fraction'.

In formula (\ref{eq:fn_class}) the presence of a non-zero superfluid fraction is related to a non-vanishing
expectation value $\langle L_z^2\rangle_q$. This allows to conclude generally that the superfluid fraction
tends to zero in the thermodynamical limit: as $L\rightarrow +\infty$, $\exp(-L {\mathcal H}_{v_{\rm rot}}/\hbar)$
becomes proportional to the projector on the ground state of the Hermitian part of the fictitious Hamiltonian.
As this ground state has a vanishing angular momentum ($l=0$), $\langle L_z^2\rangle_q$
tends to zero in the thermodynamical limit. For a non-zero $v_{\rm rot}$, one
gets similarly that $\langle P\rangle/(N m v_{\rm rot}) \rightarrow 1$ in the thermodynamic
limit.  One sees that the corresponding critical length scale is
$L \sim \hbar/\delta E$, where $\delta E$ is the energy difference between the first excited
state and the ground state of the hermitian part of ${\mathcal H}$. As the minimal energy within
a given subspace of angular moment is an increasing function of $|l|$, $\delta E$ is either
$E_{n=1,l=0}-E_{n=0,l=0}$ or $E_{n=0,l=1}-E_{n=0,l=0}$. The lengths corresponding to these
two possibilities have been identified in \cite{atom_laser}, they are respectively the correlation
length and the coherence length  of the bulk. As discussed
in \cite{atom_laser}, the coherence length is actually always larger than the correlation length.
One then sees very generally that superfluidity in the spirit of (\ref{eq:fn})
is exponentially suppressed when the length of the sample greatly exceeds the bulk coherence length.

The Bogoliubov approach in previous sections of this paper has produced a physical
picture in which a superfluid behavior can still be identified with
valleys even when the standard definition
(\ref{eq:fn}) gives a normal fraction close to unity. Within the classical field model
we can test this prediction in an exact manner, without relying on the Bogoliubov
approximation. It is useful first to identify the dimensionless parameters on which the
classical field model actually depends. Let us express the field $\psi$ in units of
$\rho^{1/2}$ (where $\rho=N/L$ is the mean density) and the spatial coordinate $z$ in units of
\begin{equation}
L_0 = \frac{\rho\hbar^2}{m k_B T}.
\end{equation}
Note that $L_0$ is on the order of the coherence length of the bulk gas,
whatever the value of $\chi$ is \cite{atom_laser}.
One then realizes that $\beta E[\psi]$, and therefore the state of the gas,
depends only on (i) $\tilde{v}_{\rm rot}$,
the velocity ${v_{\rm rot}}$ in units of $k_BT/(\rho\hbar)$, (ii) $\tilde{L}$, the length $L$ in units
of $L_0$, and (iii) on a dimensionless parameter $\chi$
controlling the interaction strength:
\begin{equation}
\tilde{v}_{\rm rot}\equiv \frac{\rho\hbar v_{\rm rot}}{k_B T} \ \ \ \ 
\tilde{L}\equiv \frac{L}{L_0}=2\pi^2 \frac{T}{T_v} \ \ \ \ 
\chi \equiv \rho L_0 \frac{\rho g}{k_B T} = \left(\frac{L_0}{\xi}\right)^2 = \left(\frac{T_{\rm df}}{T}\right)^2
\end{equation}
where $\xi$ is the healing length such that $\hbar^2/m\xi^2\equiv \rho g$,
$T_{\rm df}$ is the temperature upper bound (\ref{eq:Tdf})
required to have weak density fluctuations and $T_v$ was defined in (\ref{eq:tv})
in the Bogoliubov approach as the temperature lower bound to have several valleys populated.

\begin{figure}[htb]
\begin{center}
\includegraphics[width=10cm,clip]{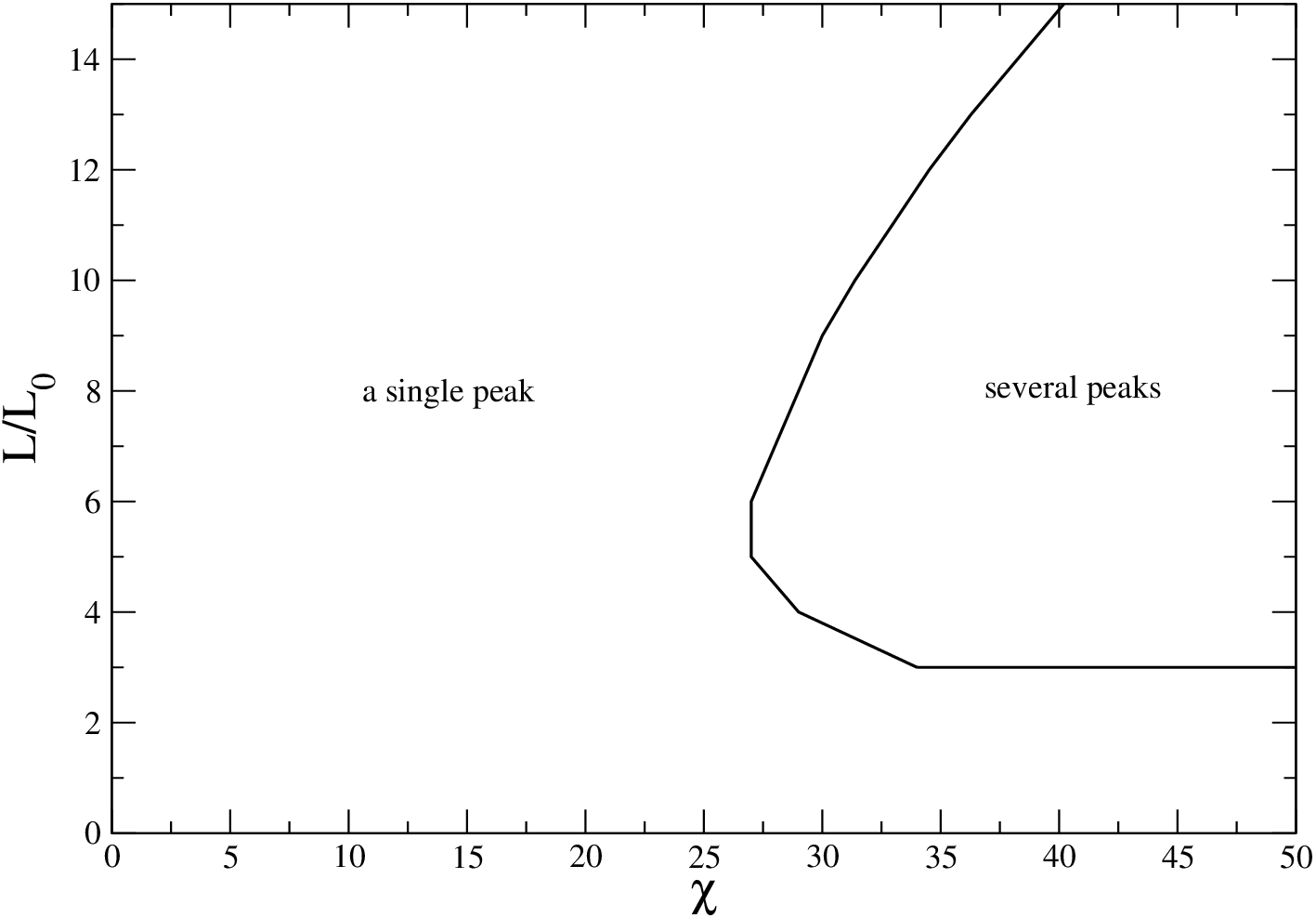}
\caption{\label{fig:phase} From a full numerical solution of the classical field model,
domain in the $\chi - \tilde{L}$ plane where the probability
distribution of $p(P)$ is multi-peaked. Only peaks higher than $10^{-3}$ times the maximal value
of $p(P)$ were considered, and $\chi$ and $\tilde{L}$ were varied in steps of one.}
\end{center}
\end{figure}

For a non-rotating gas, $v_{\rm rot}=0$, we explore the plane of the two remaining
parameters, $\tilde{L}$ and $\chi$, by a numerical diagonalization of the fictitious
Hamiltonians ${\mathcal H}_{v_{\rm rot}}$ and ${\mathcal H}_w$. This gives access to the 
probability distribution of $P$ without approximation, and allows to see in which
parameter range this distribution has several peaks. The result is plotted
in figure (\ref{fig:phase}). As expected, presence of a multi-peaked structure
requires a length $L$ larger than $L_0$, otherwise the gas is in the Bose-condensed
regime. It also requires a large enough $\chi$, that is weak enough density fluctuations.
The boundary of the multi-peaked domain is studied analytically in the next subsection,
using a large $\chi$ expansion.

For a rotating gas, we have also performed numerical diagonalizations 
of ${\mathcal H}_{v_{\rm rot}}$ and ${\mathcal H}_w$, which requires in the case of
${\mathcal H}_w$ the diagonalization of a non-Hermitian matrix. We have recovered
the phenomenon, obtained within the Bogoliubov approach, that the peaks
of a well resolved multi-peaked structure for $p(P)$ are essentially not shifted by 
the rotation of the vessel, but their amplitudes depend on $v_{\rm rot}$.
The mean momentum of the gas in the classical field model is close
to the Bogoliubov prediction in its validity domain, that is
for weak density fluctuations, see figure (\ref{fig:ptot}).

\subsection{Asymptotic expressions}
Analytical results can be obtained in two extreme cases.
First, in the ideal Bose gas case, where $\chi=0$. The fictitious Hamiltonians
appearing in (\ref{eq:gen_clas}) are then quadratic in the position and momentum
operators and can be diagonalized exactly. The difference with the usual harmonic
oscillator case is that the potential energy term $M\Omega^2(x^2+y^2)/2$ is now complex.
But one just has to choose for the ``oscillation frequency" $\Omega$ the determination
of the square root such that $\mbox{Re}\,\Omega >0$
\footnote{This is
  not possible if $M\Omega^2$ is real negative, which
can occur only for $\zeta=0$ and $\mu+m{v_{\rm rot}}^2/2 >0$. One can then still use the formulas
to come provided that one uses analytic continuation.}.
In this case the usual Gaussian wavefunction
$\propto \exp[-m\Omega(x^2+y^2)/2\hbar]$ is a perfectly normalizable ``ground state",
and the usual repeated action of the creation operators can be used to obtain
the ``excited states". As a consequence the usual 2D isotropic harmonic oscillator spectrum
is recovered, ${\mathcal E}_{n,l}=(2n+|l|+1)\Omega+im w l$ where $n$ is radial quantum number
and $l$ is the angular momentum quantum number. The characteristic function can then be
calculated exactly as the sum of a geometrical series:
\begin{equation}
g(\zeta) = \frac{\cosh(\Omega_{v_{\rm rot}} L)-\cos(m{v_{\rm rot}}L/\hbar)}{\cosh(\Omega_w L)-\cosh[mL(\zeta k_B T-i {v_{\rm rot}})/\hbar]}
\end{equation}
where the complex oscillation frequencies are such that
\begin{equation}
\Omega_w^2 = -\frac{2m}{\hbar^2} (\mu+\frac{1}{2} mw^2) 
\ \ \ \ \mbox{and}\ \ \ \
\Omega_{v_{\rm rot}}^2 = -\frac{2m}{\hbar^2} (\mu+\frac{1}{2}m{v_{\rm rot}}^2).
\end{equation}
and $w$ was introduced in \eq{w}.
The resulting $p(P)$ is shown in the appendix \ref{appen:single_peak}
to have a single maximum, at least for $v_{\rm rot}=0$.
The chemical potential was already calculated in the classical field approximation, using the Poisson
summation formula, see (\ref{eq:CFN2}).
Using (\ref{eq:P_clas}) one also recovers the expression (\ref{eq:CFP}) for the mean momentum.
So, for the ideal Bose gas in the classical field model, using the Feynman formula to relate a path
integral to the trace of an evolution operator is similar to the use of the Poisson summation formula!

\begin{figure}[htb]
\begin{center}
\includegraphics[width=10cm,clip]{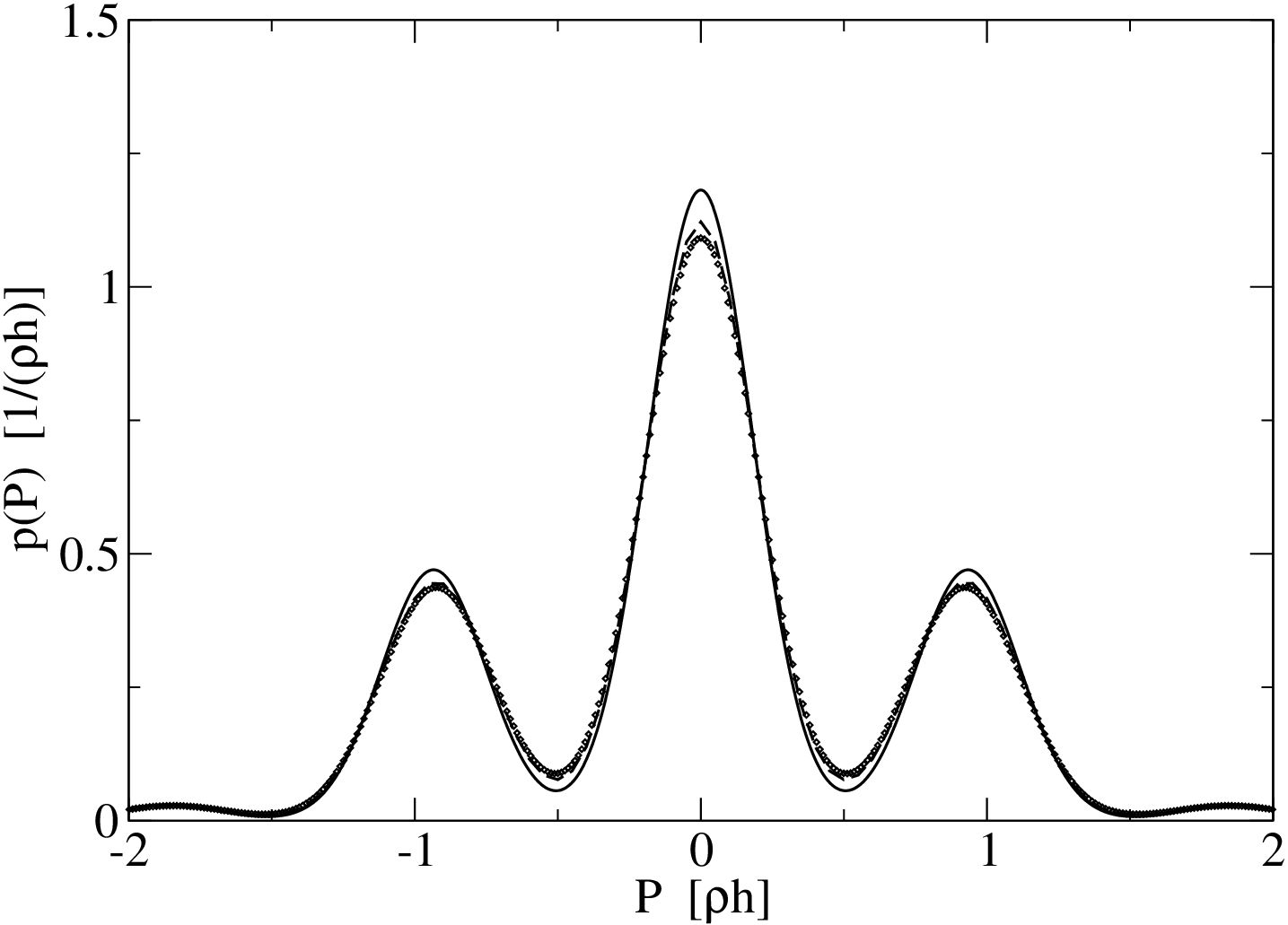}
\caption{\label{fig:comp} Probability distribution of the total momentum $P$
in the classical field model, for $\chi=200$, $v_{\rm rot}=0$ and $L/L_0=20$. Symbols: full
numerical solution. Solid line: analytical expression (\ref{eq:p_clas_analy}).
Dashed line: semi-analytical prediction resulting from the use of $r_l$ as given by the exact solution of
$U'(r_l)=0$ (see text). The dashed line and the symbols are almost indistinguishable.}
\end{center}
\end{figure}

Second, in the large $\chi$ limit, where intensity fluctuations of the
field become weak, one can obtain asymptotic formulas for $g(\zeta)$. In this limit,
the coherence length $L_0$ is much larger than the healing length $\xi$.
To simplify the calculation, we take ${v_{\rm rot}}=0$ in a first
stage and
we restrict to the case of a length $L$ on the order of a few times the coherence length and therefore
much larger than $\xi$ \cite{atom_laser}: in this case, the energy differences $E_{n=1,l}-E_{n=0,l}$
are much larger than $\hbar/L$ so that, {\sl within each subspace of fixed angular momentum} $l$,
one can restrict to the ground state of the Hermitian part of the fictitious Hamiltonian
in the calculation of $g(\zeta)$:
\begin{equation}
g(\zeta) \propto \sum_{l\in\mathbb{Z}} e^{-L(E_{0,l}+ml\zeta k_B T)/\hbar}
\label{eq:g_ground}
\end{equation}
where the normalization factor is obtained from $g(0)=1$ and $E_{0,l}$ is the lowest eigenenergy
of the Hermitian part of ${\mathcal H}_w$ with angular momentum $l$ (remember that $v_{\rm rot}=0$ here).
Using polar coordinates $r,\theta$, we write the corresponding wavefunction
$\phi_{0,l}$ as
\begin{equation}
\phi_{0,l}(x,y) = f_l(r) \frac{e^{il\theta}}{(2\pi r)^{1/2}}
\end{equation}
where $r=(x^2+y^2)^{1/2}/\rho^{1/2}$.
The purely radial wavefunction then solves the Schr\"odinger equation
\begin{equation}
-\frac{1}{2} \frac{d^2}{dr^2}f_l(r)  + U_l(r) f_l(r) = \tilde{E}_{0,l} f_l(r)
\end{equation}
with an effective potential $U_l$ including a centrifugal term:
\begin{equation}
U_l(r)=\frac{l^2-1/4}{2 r^2} +\frac{1}{2}\chi r^4 -(\tilde{\mu}-\tilde{\zeta}^2/2) r^2
\end{equation}
where $\tilde{\zeta}=\rho\hbar\zeta$, $\tilde{E}_{0,l}=L_0 E_{0,l}/\hbar$, and where
the reduced chemical potential $\tilde{\mu}=\rho L_0\mu/(k_B T)$ is here close
to its bulk value calculated for large $\chi$ in \cite{atom_laser}, since $L$ exceeds a few $L_0$:
\begin{equation}
\tilde{\mu} = \chi + \frac{1}{2} \chi^{1/2} + O(1).
\end{equation}
In the large $\chi$ limit, the ground state is
deeply localized in the minimum of $U_l(r)$ occurring at a non-zero distance $r_l$ from
the origin, solution of $U'(r_l)=0$. One can then expand $U_l(r)$ in a power series around $r=r_l$, include
the quadratic part in $(r-r_l)^2$ in a harmonic oscillator diagonalization, include
the cubic, quartic, $\ldots$ terms with perturbation theory. Treatment of up to quartic
terms with second order perturbation theory turns out to be sufficient here:
\begin{equation}
\tilde{E}_{0,l} = U_l(r_l) + \frac{1}{2}\omega_l  + \frac{3 U_l^{(4)}(r_l)}{96\omega_l^2}
-\frac{11 \left[U_l^{(3)}\right]^2}{288\omega_l^4} + \ldots
\label{eq:E0_clas}
\end{equation}
where $\omega_l=[U^{(2)}(r_l)]^{1/2}$ is the oscillation frequency in the harmonic
approximation to $U_l(r)$.
The position $r_l$ can be calculated exactly, since $r_l^2$ is found to be the root of
a cubic equation; when (\ref{eq:E0_clas}) is plugged into (\ref{eq:g_ground}), one gets a very
good approximation for $g(\zeta)$ and, by numerical Fourier transform, a very good approximation
for the probability distribution of $P$, as compared to the full numerical solution, for $\chi\gg 1$,
see figure~\ref{fig:comp}. A more tractable expression can be obtained by realizing that typical values of
$\tilde{\zeta}^2$ and $l^2$ are of order $\chi^{1/2}$ \footnote{This
  can be obtained by trial and error, but also by the fact that,
in the Bogoliubov theory, the narrowest structure in the probability distribution of $P$
scales as $(f_n^v)^{1/2}$, resulting in a maximal value of $\zeta$ scaling as $1/(f_n^v)^{1/2}$.
In the temperature regime $k_B T > \mu$, which is the one of the classical field model,
$f_n^{v,0}$ scales as $1/\chi^{1/2}$, see (\ref{eq:f_nHighT}).},
 and by performing a systematic
expansion of the cubic equation for $r_l$ and of (\ref{eq:E0_clas}) in powers of ${\chi}^{-1/2}$:
\begin{equation}
\tilde{E}_{0,l}-l \tilde{\zeta} = \mbox{const}+
\frac{1}{2}(\tilde{z}-l)^2+\frac{l^2}{2\chi^{1/2}}-\frac{1}{8\chi}(l^2-\tilde{\zeta}^2)^2
+O(\chi^{-1/2})
\label{eq:E0_analy}
\end{equation}
where $\mbox{const}$ depends on $\chi$ only, not on $l$ or $\zeta$. The second term in the righthand
side of the equation is $O(\chi^{1/2})$, the third and fourth terms are {\sl a priori} of the same
order $O(1)$. In practice the fourth term is typically $O(\chi^{-1/2})$ so we neglect it
\footnote{After multiplication of (\ref{eq:E0_analy}) by $-L/L_0$ and exponentiation
to get $g(\zeta)$, one realises that $g(\zeta)$, considered as a function of $\tilde{\zeta}$,
is approximately a superposition of narrow peaks centered
in integer values $l$ and of width $\propto (L_0/L)^{1/2}$, with an envelope of width $\chi^{1/4}(L_0/L)^{1/2}$.
As a consequence, for a given $l$, $\tilde{\zeta}-l$ is $O((L_0/L)^{1/2})$, and the fourth term
in the RHS of (\ref{eq:E0_analy}) is $O(l^2/\chi)=O(\chi^{-1/2})$.}.
We then obtain a Gaussian expressions for $g(\zeta)$ and $p(P)$.

The previous calculations are immediately generalized to the case of a non-zero velocity: as
$L>L_0$, $\mu+m{v_{\rm rot}}^2/2$ depends weakly on ${v_{\rm rot}}$ and can be replaced by the bulk value for ${v_{\rm rot}}=0$;
one then has to replace $\tilde{\zeta}^2$ by $\tilde{\zeta}^2-2i\tilde{v}_{\rm rot}\tilde{\zeta}$
in the above calculation of $\tilde{E}_{0,l}$, where $\tilde{v}_{\rm rot}$, being independent on $\chi$, is $O(1)$.
We then obtain for $g(\zeta)$:
\begin{equation}
g(\zeta) = A \sum_{l\in \mathbb{Z}} e^{-\tilde{L}(l-\tilde{\zeta}+i\tilde{v}_{\rm rot})^2/2}e^{-\tilde{L}l^2/(2\chi^{1/2})}
\label{eq:toto}
\end{equation}
where $A$ is a constant factor such that $g(0)=1$.
Performing the Fourier transform of this expression and using the Poisson summation formula leads to
\begin{equation}
p(P) = B e^{-(\tilde{P}-\tilde{L}\tilde{v}_{\rm rot})^2/(2\tilde{L})}
\sum_{q \in {\mathbb Z}} e^{-(\tilde{P}-2\pi q)^2\chi^{1/2}/(2\tilde{L})}
\label{eq:p_clas_analy}
\end{equation}
where $\tilde{P}=P/(\rho\hbar)$. The constant factor $B$ is such that the integral
of $p(P)$ over $P$ is equal to unity. Formula (\ref{eq:p_clas_analy}) is very suggestive,
as it is simply a Gaussian envelope centered in $\tilde{P}=\tilde{L}\tilde{v}_{\rm rot}$
on top of a periodic train of Gaussian peaks separated by $2\pi$ in $\tilde{P}$
space. Therefore the function $p(P)$ is multi-peaked if the envelope 
has a width larger than $2\pi$ and if each Gaussian of the train has a width less than $2\pi$:
\begin{equation}
\tilde{L}^{1/2} > 2\pi \ \ \ \mbox{and} \ \ \ (\chi^{1/2}/\tilde{L})^{1/2} > 2\pi.
\end{equation}
One recovers the conditions (\ref{eq:regime}) using the fact that
$f_n^0=\chi^{-1/2}$ 
in the classical field regime $k_B T > \rho g$, see (\ref{eq:f_nHighT}).

The asymptotically exact expression (\ref{eq:p_clas_analy})
is successfully compared with the full numerical solution
in figure~\ref{fig:comp}. One can also compare it to the Bogoliubov prediction (\ref{eq:p_PBog}):
the two predictions are found to be identical up to higher order terms (see appendix \ref{appen:cc_vs_Bog}).
As a consequence, all the physical discussion following (\ref{eq:p_PBog}) also holds for the
classical field model!
 From (\ref{eq:toto})
one can also calculate the normal fraction $f_n^0$ for ${v_{\rm rot}}=0$ by taking
the second order derivative of (\ref{eq:toto}) with respect to $\zeta$. Equivalently one may calculate the
variance of $P$ from (\ref{eq:p_clas_analy}) and use (\ref{eq:fn0}). One gets two equivalent forms:
\begin{equation}
f_n^0 = 1 -\frac{L}{L_0} \frac{\sum_q q^2 \exp(-\tilde{L}_{\rm eff}q^2/2)}{\sum_q \exp(-\tilde{L}_{\rm eff}q^2/2)}
= \frac{1}{{\chi}^{1/2}+1} + \frac{1}{L(1+\chi^{-1/2})^2} \frac
{\sum_{q\in {\mathbb Z}} (2\pi q)^2 e^{-(2\pi q)^2/(2 \tilde{L}_{\rm eff})}}
{\sum_{q\in {\mathbb Z}} e^{-(2\pi q)^2/(2 \tilde{L}_{\rm eff})}}
\end{equation}
where the reduced effective length is $\tilde{L}_{\rm eff}=L(1+\chi^{-1/2})/L_0$. The first form immediately
shows that $f_n^0$ tends to one exponentially for $L\gg L_0$. The second form recovers the formula (11)
of \cite{Svistunov} in a calculation up to first order in ${\chi}^{-1/2}$ where one replaces
$\rho_S$ of \cite{Svistunov} by $\rho(1-{\chi}^{-1/2})$.

\section{Conclusions}

We have investigated the superfluid properties of a ring of degenerate
and weakly interacting 1D Bose gas at thermal equilibrium with a
rotating vessel at velocity $v_{\rm rot}$. Provided the transverse
trapping is strong enough, our model is a good description of a Bose
gas confined in a toroidal trap~\cite{torus}.

Using the conventional definition of the superfluid fraction,
which relies on the variance of the total momentum of the gas in the
limit $v_{\rm rot}\rightarrow 0$,
we find that the gas has a significant superfluid fraction only in the
Bose condensed regime, that is when the length of the ring does not
exceed the coherence length of the bulk gas $l_c \sim \rho \lambda^2$,
$\rho$ being the mean density and $\lambda$ the thermal de Broglie
wavelength.

To investigate more carefully the regime where the length of the ring
exceeds the coherence length, we have 
considered the full probability distribution of the total momentum $P$. We have identified a regime
where several peaks appear in this probability distribution, each peak corresponding to a quasicondensate
in a plane wave state with a given winding number, the analog of supercurrents in superconductors.
Each supercurrent state exhibits some superfluid behaviour: in presence of a non-zero
$v_{\rm rot}$, the peaks in the probability distribution of $P$ are
indeed not shifted. This allows to define
a local normal fraction for an individual supercurrent.
Quantitatively, we have found that the probability
distribution of $P$ shows several isolated peaks provided that
the length of the ring in units of the coherence length does not
exceed the inverse of the local normal fraction.

In this non-Bose-condensed regime, it is obvious that the conventional
criterion for superfluidity based on the variance of $P$ is sensitive
to the envelope of the distribution but does not catch its
multi-peaked structure.
To get it, a direct measurement of the total current $P$ is
required, which, e.g., could be performed by means of the technique
proposed in~\cite{artoni}: 
in a slow-light regime, the dielectric susceptibility of the atoms
depends on the local value of the matter current so that the
phase accumulated by light after a round-trip around the ring is proportional
to the total current $P$.

\appendix

\section{Absence of multiple peaks in $p(P)$ for the ideal gas}
\label{appen:single_peak}

The characteristic function
for the ideal gas in the classical field regime is:
\begin{equation}
\eqname{g_P0cl}
g(\zeta)=\prod_k\frac{\beta(\epsilon_k-\mu)}{\beta(\epsilon_k-\mu)-i\zeta
  \hbar k}.
\end{equation}
In the $v_{\rm rot}=0$ case, we can regroup the pairs $\pm k$ and
rewrite \eq{g_P0cl} as:
\begin{equation}
\eqname{g_P0cl1}
g(\zeta)=\prod_{k>0}g_k(\zeta)=
\prod_{k>0}\frac{\beta^2(\epsilon_k-\mu)^2}{\beta^2(\epsilon_k-\mu)^2+(\hbar k)^2\,\zeta^2}.
\end{equation}
The Fourier transform $p_k(P)$ of each term $g_k(\zeta)$ is then:
\begin{equation}
p_k(P)=\frac{\eta_k}{2}e^{-\eta_k\,|P|},
\end{equation}
with $\eta_k=\beta(\epsilon_k-\mu)/\hbar k$. In particular, $p_k(P)$
has the property of being an even function that is decreasing for $P>0$
(let's call this property ${\mathcal P}$).
As the characteristic function $g(\zeta)$ is the product of the $g_k(\zeta)$, the
distribution function $p(P)$ is the convolution of the $p_k(P)$.
As the convolution of two functions with the property ${\mathcal P}$
gives again a function with the property ${\mathcal P}$ (a sketch of
the proof is given below), we can conclude that the
probability distribution $p(P)$ for the ideal gas has a single
maximum, which is at $P=0$. The possibility of multi-peaked
structures is therefore ruled out for the ideal gas in the classical
field regime.

We can prove that the property ${\mathcal P}$ is preserved by
convolution operations in the following way. Let $p_1(P)$ and $p_2(P)$
be two arbitrary functions sharing property ${\mathcal P}$.
We have to prove that:
\begin{equation}
p_c(P)=\int\!dP'\,p_1(P')\,p_2(P-P')
\eqname{convol}
\end{equation}
\begin{itemize}
\item[i)] is an even function.

\item[ii)] is a monotonically decreasing function for $P>0$.
Let's compute its derivative
  for $P>0$:
\begin{multline}
p_c'(P)=\int\!dP'\,p_1(P')\,p_2'(P-P')=\int\!dP'\,p_1(P-P')\,p_2'(P')=\\
=\int_0^\infty\!dP'\,p_2'(P')\Big[p_1(P-P')-p_1(P+P')\Big].
\end{multline}
As $|P-P'|<|P+P'|$ and $p_1$ is a decreasing function of the absolute
value of its argument, the integrand is negative. This guarantees that
$p_c'(P)<0$ for all $P>0$.
\end{itemize}

\section{Derivation of the Bogoliubov Hamiltonian and momentum operator}
\label{appen:Bog}

The Hamiltonian \eq{HBog}
can be obtained either by directly solving the Bogoliubov-de Gennes
equations for a moving system, or, better, by applying Galilean
invariance arguments to the well-known case of a system at rest.
The eigenstates of a weakly interacting Bose gas at rest ($v_0=0$)
are labeled within Bogoliubov theory by the occupation
number of the bosonic quasiparticle modes $\{n_k\}$.
Thanks to translational invariance, these eigenstates are also
eigenstates of the momentum: each quasiparticle carrying a momentum
$\hbar k$, the total momentum of the gas is given by:
\begin{equation}
P[\{n_k\}]=\sum_{k\neq0} \hbar k\,n_k.
\end{equation}
Omitting for the moment the rotation energy $-P  v_{\rm
  rot}$, the total energy is given by:
\begin{equation}
E_{v_{\rm rot}=0}[\{n_k\}]=E_{GP}[v_{\rm rot}=0,v_0=0]-\sum_{k\neq0} \epsilon^0_k\,V_k^2+\sum_{k\neq0}
\epsilon^0_k\,n_k.
\end{equation}

The properties of a moving quasi-condensate at $v_0\neq 0$ can be obtained
from the ones of a quasi-condensate at rest by transforming the energy and
the momentum via a Galilean transformation of velocity $v=v_0$.
As discussed in sec.\ref{sec:model}, the total momentum $P'$ in the
moving frame is in fact given by:
\begin{equation}
P'[\{n_k\}]=P[\{n_k\}]+N m v_0=\sum_{k\neq0} \hbar k\,n_k+N m v_0.
\eqname{P'}
\end{equation}
Inserting back the rotation term $-P' v_{\rm rot}$, the energy $E'$ turns
out to be:
\begin{multline}
E'[\{n_k\}]=E_{v_{\rm rot}=0}[\{n_k\}]+\frac{1}{2}Nm v_0^2+(v_0
-v_{\rm rot}) P[\{n_k\}]=\\ =E_{GP}[v_{\rm rot},v_0]-\sum_{k\neq0} \epsilon^0_k\,
V_k^2+\sum_{k\neq0}\big[\epsilon^0_k+\hbar k (v_0-v_{\rm rot})\big]\,n_k.
\eqname{E'}
\end{multline}
The eigenstates and eigenenergies obtained in this way exactly
correspond to the ones of the Bogoliubov Hamiltonian \eq{HBog}.
For each state, the total momentum \eq{P'} agrees with \eq{Pjoli}.

\section{Comparison of Bogoliubov and classical field theory for $p(P)$}
\label{appen:cc_vs_Bog}
In the regime of weak density fluctuations and a temperature $k_B T \gg \rho g$
we compare the expressions for the total momentum probability distribution $p(P)$
obtained by Bogoliubov theory, (\ref{eq:p_PBog}) on one side, and by a large $\chi$ expansion
of the classical field model, (\ref{eq:p_clas_analy}) on the other side.
At first glance, (\ref{eq:p_clas_analy}) looks much simpler and therefore different than
(\ref{eq:p_PBog}). However one has the identity
\begin{equation}
\tilde{L}^{-1}(\tilde{P}-\tilde{L}\tilde{v}_{\rm rot})^2+\tilde{L}^{-1}(\tilde{P}-2\pi q)^2\chi^{1/2} =
\tilde{L}^{-1}(1+\chi^{1/2})\left(\tilde{P}-\tilde{P_0}\right)^2
+\tilde{L}^{-1}\frac{(2\pi q-\tilde{L}\tilde{v}_{\rm rot})^2}{1+\chi^{-1/2}}
\label{eq:ident}
\end{equation}
where
\begin{equation}
\tilde{P}_0=\frac{\tilde{L}\tilde{v}_{\rm rot}+2\pi q{\chi}^{1/2}}{{\chi}^{1/2}+1}.
\end{equation}
Expanding ${\tilde P}_0$ up to first order in ${\chi}^{1/2}$ and using the fact that $f_n^{v,0}={\chi}^{-1/2}$
in the Bogoliubov theory, one finds that $\tilde{P}-\tilde{P_0}$ coincides with
the expression in square brackets in (\ref{eq:p_PBog}). The factor in front of this expression,
proportional to $1/f_n^{v,0}$, recovers ${\tilde
  L}^{-1}(1+\chi^{1/2})$ within leading order in ${\chi}^{1/2}$.
The last term in (\ref{eq:ident}) coincides with the argument of the first exponential
factor in (\ref{eq:p_PBog}) when expanded up to first order in ${\chi}^{-1/2}$.

\section*{Acknowledgements}

We acknowledge useful discussions with Anthony Leggett, Jean Dalibard, Lev Pitaevskii.
I.C. acknowledges a Marie Curie grant from the EU under contract number
HPMF-CT-2000-00901.
Laboratoire Kastler Brossel is a Unit\'e de
Recherche de l'\'Ecole Normale Sup\'erieure et de l'Universit\'e Paris
6, associ\'ee au CNRS.


\begin{thebibliography}{00}
\bibitem{chemi} We used the fact that the chemical potential $\mu$ varies
only to second order in $v_{\rm rot}$ when the mean density is fixed.
\end{thebibliography}
\end{document}